\documentclass[prb,amsmath,amssymb,twocolumn,lengthcheck,showpacs,floatfix,groupedaddress,superscriptaddress]{revtex4}
\usepackage{graphicx}
\usepackage[english]{babel}
\usepackage{times}
\usepackage{dcolumn}
\usepackage[usenames,dvipsnames]{color}
\usepackage{units}
\usepackage[usenames,dvipsnames]{color}
\usepackage{bm}

\usepackage{color}

\newcommand{\vc}[1]{{\mathbf{#1}}}

\renewcommand{\bm}[1]{\mathbf{#1}}

\begin{document}

\title{Fine structure of the spectra of the Kondo lattice model: two-site cellular DMFT study}

\author{\v{Z}iga Osolin}

\affiliation{Jo\v{z}ef Stefan Institute, Jamova 39, SI-1000 Ljubljana,
Slovenia}

\author{Rok \v{Z}itko}

\affiliation{Jo\v{z}ef Stefan Institute, Jamova 39, SI-1000 Ljubljana,
Slovenia}
\affiliation{Faculty  of Mathematics and Physics, University of Ljubljana,
Jadranska 19, SI-1000 Ljubljana, Slovenia}

\date{\today}

\begin{abstract}
We study the antiferromagnetic and the paramagnetic Kondo insulator phases of the Kondo lattice model on the cubic lattice at half-filling using the cellular dynamical mean-field theory (CDMFT) with numerical renormalization group (NRG) as the impurity solver,
focusing on the fine details of the spectral function and self-energy. We find that the non-local correlations increase the gap in both the antiferromagnetic and the Kondo insulator phase and shrink the extent of the antiferromagnetic phase in the phase diagram but do not alter any properties qualitatively. The agreement between the numerical CDMFT results and those within a simple hybridization picture, which adequately describes the overall band structure of the system but neglects all effects on the inelastic-scattering processes, is similar to that of the single-site DMFT results; there are deviations that are responsible for the additional fine structure, in particular for the asymmetric spectral resonances or dips that become more pronounced in the strong-coupling regime close to the antiferromagnet-paramagnetic quantum phase transition. These features appear broader in the CDMFT mostly due to numerical artifacts linked to more aggressive state truncation required in the NRG.

\end{abstract}

\pacs{71.27.+a, 72.15.Qm, 75.20.Hr, 75.30.Mb}

\maketitle

\section{Introduction}
Heavy fermion materials have unusual properties due to strong correlations that still lack a complete microscopical understanding despite many years of continuous research \cite{hewson1997kondo,RevModPhys.56.755}. These compound materials consist of lanthanide or actinide atoms and other metallic elements. Typical examples are $\mathrm{Ce_3Bi_4Pt_3}$, $\mathrm{YbB_{12}}$, $\mathrm{CeNiSn}$, $\mathrm{SmB_6}$, and $\mathrm{CeRh_2Si_2}$ \cite{jaime2000closing,sugiyama1988field,PhysRevB.81.094403,andres19754}. The correlation in these materials is due to localized $f$ orbitals with strong on-site repulsion that are coupled to conduction-band electrons (spd bands), resulting in high effective mass enhancement of their Fermi-liquid quasi-particles. Consequently, phenomena such as unconventional (spin-mediated) superconductivity, complex magnetism, huge thermopower, and, in general, very rich phase diagrams are found in heavy fermions \cite{mathur1998magnetically,bauer2004heavy}. In some cases, these materials have semiconducting or insulating properties at low temperatures (Kondo insulators) \cite{cooley1995sm,mason1992spin,iga1998single}.

The minimal model for heavy fermions is the Kondo lattice model (KLM) \cite{Kasuya01071956}. It consists of two bands: a non-interacting itinerant band (spd band) coupled to a localized orbital at each lattice site (f band). Projecting out high-energy double and zero occupancy f states through the Schrieffer-Wollf transformation allows us to represent the f electrons as local moments coupled to the itinerant electrons with coupling constant $J$ \cite{schrieffer1966relation}. The KLM qualitatively describes the crucial features of heavy fermion materials. At high temperatures, the f moments are nearly free since the itinerant electrons are effectively decoupled, thus the material behaves as a conventional metal. At low temperatures, a local f moment is screened by itinerant electrons (Kondo effect) \cite{Kondo01071964,RevModPhys.47.773}. The lattice variant, where the screening occurs at each site, leads to a coherent state which is a strongly renormalized Fermi liquid with f states included in the Fermi volume; the f states thus become itinerant as well. Exactly at half-filling, the chemical potential lies inside the gap between the resulting effective bands and the system is insulating, while at finite doping the chemical potential lies in a part of the band with very flat dispersion, giving rise to the heavy-fermion behavior.


In this work we study the KLM on a cubic lattice exactly at half-filling. There are two competing effects \cite{doniach1977kondo}: the effective RKKY\cite{ruderman1954indirect} interaction stabilizes the antiferromagnetic insulator for $J < J_c$, while at higher $J$, the system is a paramagnetic Kondo insulator.  
In our previous single-site DMFT study in Ref. \onlinecite{PhysRevB.91.075105}, we have found additional fine structure of spectra  in the antiferromagnetic phase. In the momentum-resolved spectral functions, we have observed that the hybridized bands are not truly degenerate at the band center and that the local (momentum-integrated) spectral function exhibits narrow features, ``spin resonances'', inside the bands. They become more pronounced in the strong-coupling Kondo antiferromagnet, where they can be easily distinguished from the gap edges. The origin of spin resonances can be explained as follows. In the paramagnetic phase, a single-pole hybridization ansatz for the self-energy is valid. In the antiferromagnetic phase, we need to extend it to account for the staggered magnetizations of itinerant and $f$ moments as \cite{hoshino2010}
\begin{equation}
\label{eq:ansatz}
\Sigma_{\alpha \sigma}(z) = \alpha \sigma h + \frac{\tilde{V}^2}{z-\alpha \sigma H},
\end{equation}
where $\alpha=+1,-1$ corresponds to the sublattice A or B, $\sigma$ is the spin, $\tilde{V}$ can be interpreted as the renormalized hybridization
between the c and f bands in the hybridization picture for the
periodic Anderson model, $h$ is the staggerer exchange field of band electrons, and $H$ is the staggered exchange field of f electrons.
If the quasi-local compensation criterion, $h=-H$ \cite{hoshino2010} is slightly violated, there is an avoided crossing of branches around 
\begin{equation}
\label{eq:omegaStar}
\omega^* = \sqrt{\tilde{V}^2+h^2}.
\end{equation}
That would lead to an opening of a small gap in a non-interacting system. Because the system is interacting, however, the imaginary part of the self-energy is non-zero, and the avoided crossing results in the spectral function amplification (spin resonance) or reduction (dip) around $\omega^*$.

There have been concerns that these features are an artifact of the single-site approximation of the DMFT and that they would not pesist in a more refined treatment of the problem. The inclusion of non-local fluctuations can change the structure and phases, an example of which is the pseudo-gap formation in the Hubbard model \cite{PhysRevB.73.165114,PhysRevB.76.045108,huscroft2001pseudogaps}.

Existing two-site CDMFT studies using quantum Monte Carlo (QMC) impurity solver on the related periodic Anderson model mostly focus on details away from half-filling (heavy fermion metallic phase), where they find that the antiferromagnetic N\'eel temperature is significantly lowered compared to the DMFT results \cite{PhysRevB.84.115105,PhysRevLett.95.016402}. The dual-fermion study in Ref. \onlinecite{li2013kondo}, however, suggests that the antiferromagnetic phase is not shrunk significantly, so long-range correlation appear not to play an important role in this problem. 

This work extends the single-site DMFT study by the cellular DMFT (CDMFT) calculation with the numerical renormalization group (NRG) as an impurity solver, thus also partially taking into account the non-local fluctuations. We study the transition from the weak coupling (band/Slater antiferromagnet) via the strong-coupling (Kondo antiferromagnet) regime to the paramagnetic Kondo insulator at exactly half-filling, focusing on the detailed structure of the spectral functions accessible with the CDMFT(NRG). The method works on the real axis so there is no need for analytical continuation as in the QMC, where it introduces new uncertainties to the spectral function calculation.

The paper is organized as follows. In Sec. \ref{sec:method}, we describe the KLM and the CDMFT(NRG) method. In Sec. \ref{sec:trunc}, we discuss the systematic errors in the CDMFT(NRG). In Sec. \ref{sec:main}, the comparison between the single-site DMFT and two-site CDMFT is presented, focusing on the dependence of the spin resonance, gap size, and phase transition. In Appendix \ref{sec:appendix}, we discuss the discretization of the impurity problem, followed by the construction of the Wilson chain for multi-channel impurity model in Appendix \ref{sec:wilsonChain2}.

\section{Model and method}
\label{sec:method}
The KLM consists of a lattice of local moments $\bm{s}^f_i$ coupled to the conduction band electrons through the on-site exchange coupling $J>0$, 
\begin{equation}
H = \sum_{\bm{k} \sigma} \epsilon_\bm{k} c_{\bm{k} \sigma}^\dagger c_{\bm{k} \sigma} + J \sum_i \bm{s}^f_i \cdot{} \bm{s}^c_i,
\end{equation}
where $c_{i\sigma}$ annihilates the conduction electron at site $i$, $c_{k\sigma}$ is the corresponding Fourier transform, $\bm{s}^c_i=\frac{1}{2}\sum_{\sigma \sigma'} c_{i\sigma}^\dagger \left( \bm{\tau}_{\sigma \sigma'} \right) c_{i\sigma'}$ (with $\bm{\tau}$ as Pauli matrices) is the spin of the conduction electron at site $i$, and $\epsilon_{\bm{k}}$ is the dispersion relation for the non-interacting conduction-band (c) states. For the cubic lattice, the dispersion relation reads
\begin{equation}
\epsilon_{\bm{k}} = -2t \left( \cos{k_1} + \cos{k_2} + \cos{k_3} \right)
\end{equation}
for lattice spacing $a=1$ and nearest-neighbor hopping $t$.

In the cellular DMFT, we choose a super-cell that tiles the original lattice. In our case, the super-cell contains two sites, $\alpha=A,B$, and supports N\'eel antiferromagnetic order. The non-interacting Green's function in the super-cell basis for the cubic lattice is thus
\begin{equation}
G_0^{-1}(\bm{K},z) = \begin{pmatrix}
(z + \mu) \bm{I} & -\xi_\bm{K} \bm{I} \\
-\xi^*_{\bm{K}} \bm{I} & (z + \mu) \bm{I}
\end{pmatrix},
\end{equation}
where $z=\omega + i \eta$ ($z = \omega - i\eta$) corresponds to retarded (advanced) Green's function $G_0(\bm{K},z)$ with $\eta>0$ infinitely small, the matrix blocks correspond to A and B sites, internal block indexes are used for spins, and
\begin{align}
\xi_{\bm{K}} =& -t \left(1 + e^{i(K_1 + K_2)} + e^{iK_1} \right. \nonumber \\
& \left. + e^{iK_2} + e^{iK_3} + e^{i(K_1+K_2-K_3)}\right),
\end{align}
where $K_l = \bm{K}\cdot{}\bm{A}_l$ are values in range $[-\pi,\pi)$, with $\bm{A}_l$ vectors that define the super-cell:
\begin{align}
\bm{A}_1 = a (1,1,0) \nonumber \\
\bm{A}_2 = a (1,-1,0) \nonumber \\
\bm{A}_3 = a (1,0,1). 
\end{align}

In the CDMFT, the self-energy is assumed to be of the specific form
\begin{equation}
\label{eq:selfEnApprox}
\Sigma_{\alpha \alpha' \sigma}(\bm{R},\bm{R}', z) = \Sigma_{\alpha,\alpha',\sigma}(z) \delta_{\bm{R},\bm{R}'}.
\end{equation}
In other words, the self-energy is translation invariant in the super-lattice, and describes the interactions within the super-cells exactly. The interactions between the super-cells are accounted for in a mean field way. The $\bm{K}$-resolved lattice Green's function in the CDMFT is
\begin{equation}
\bm{G}^{-1}(\bm{K},z) = \begin{pmatrix}
(z + \mu) \bm{I} - \bm{\Sigma}_{AA}(z) & -\xi_\bm{K} \bm{I} - \bm{\Sigma}_{AB}(z) \\
-\xi^*_{\bm{K}} \bm{I} - \bm{\Sigma}_{BA}(z) & (z + \mu) \bm{I} - \bm{\Sigma}_{BB}(z)
\end{pmatrix},
\end{equation}
Using the self-energy approximation in Eq. \eqref{eq:selfEnApprox}, one can transform the lattice problem to an impurity problem subject to the self-consistency equation \cite{kotliar2001cellular,RevModPhys.77.1027,RevModPhys.68.13}
\begin{equation}
\label{eq:selfConsistency}
\bm{\Delta}(z) = (z+\mu)\bm{I} - \bm{t} - \bm{G}_{\mathrm{local} }^{-1}(z) - \bm{\Sigma}(z), 
\end{equation}
with
\begin{equation}
\bm{t} = \begin{pmatrix} 0 & -t\bm{I} \\ -t\bm{I} & 0 \end{pmatrix}
\end{equation}
and
\begin{equation}
\label{eq:Glocal}
\bm{G}_{\mathrm{local} }(z) = \frac{1}{N} \sum_\bm{K} \bm{G}(\bm{K},z),
\end{equation}
where $N$ is the number of super-cells in the lattice.

The impurity of the impurity problem contains sites within one super-cell
\begin{align}
H_{imp} =& \sum_{\alpha,\sigma} \epsilon_{0,\alpha} d_{\alpha,\sigma}^\dagger d_{\alpha,\sigma} - t \sum_{\sigma} \left( d_{A,\sigma}^\dagger d_{B,\sigma} + H.c. \right) + \nonumber \\
& J \sum_{\alpha} \bm{s}^d_{\alpha,i} \cdot \bm{s}^f_{\alpha,i},
\end{align}
where $d_{\alpha,\sigma}$ is an annihilation operator at the impurity, $\epsilon_{0,\alpha}$ are the on-site energies of the impurity, $\alpha$ is the site index within the super-cell, and $\sigma$ is the spin index. The coupling to the bath and the bath energies are completely characterized by the hybridization function $\bm{\Delta}(z)$, that in the two-site cluster case takes the form of a $4 \times 4$ matrix.

The self-consistency is achieved by an iterative scheme: we start with a hybridization function $\bm{\Delta}(z)$, solve the impurity problem to obtain the self-energy followed by the self-consistency Eq. \eqref{eq:selfConsistency} to obtain a new $\bm{\Delta}(z)$ until the integrated difference between consecutive local Green's functions is less than $\epsilon$, in our case $\epsilon = 10^{-4}$. 

We use the NRG to solve the resulting impurity problem \cite{RevModPhys.47.773}. Because the NRG works on the real frequency axis, the spectral representation of the hybridization $\bm{\Delta}(z)$ is needed. For the CDMFT, the hybridization function has off-diagonal matrix elements and both advanced and retarded hybridization is needed to find its spectral representation:
\begin{equation}
\label{eq:gammaRealAxis}
\bm{\Gamma}(\omega) = \frac{i}{2} \left[ \bm{\Delta}(\omega+i\eta) - \bm{\Delta}(\omega-i\eta) \right]. 
\end{equation}
The bath is represented by a set of discrete states corresponding to logarithmic intervals $\pm{}[ \Lambda^{1-n-z-1}, \Lambda^{1-n-z} ]$, where $\Lambda>1$ is the discretization parameter. Every interval couples to the impurity (star basis) and $z \in (0,1]$ defines the interleaved meshes \cite{PhysRevB.49.11986}.  The star representation is then transformed to the semi-infinite Wilson chain. The calculation of the coefficients in the Wilson chain from the hybridization function is given in Appendix. The impurity model has two channels corresponding to A and B sites with non-zero inter-chain coefficients on the Wilson chain. The off-diagonal coefficients can be complex.

The chain is iteratively diagonalized,  keeping only the low-energy states. The truncation is controlled by the number of states kept, $N_{\mathrm{keep} }$ or the highest-energy state kept, $E < E_{\mathrm{cutoff}} \Lambda^{-(n+1)/2}$ for the $n$-th iteration. We use full density matrix to calculate the spectra \cite{PhysRevLett.99.076402,PhysRevLett.85.1508}. We make use of charge and spin $z$ projection symmetry of the problem.

The self-energy matrix is calculated using the correlator \cite{1998JPCM...10.8365B}
\begin{equation}
F_{ij}(\omega) = i \int_{0}^{\infty} e^{i\omega{}t} \langle \left\{ \left[d_i, H_{imp} \right](t), d_j^{\dagger} \right\} \rangle \mathrm{dt}.  
\end{equation}
In the NRG, we calculate its spectral representation, $\bm{A}_F(\omega)$. One then calculates retarded and advanced correlator,
\begin{align}
\mathrm{Re} \bm{F}(\omega \pm i\eta) = P \int \frac{\mathrm{Re}A_F(\omega')d\omega'}{\omega - \omega'} \pm \pi \mathrm{Im}A_F(\omega), \nonumber \\
\mathrm{Im} \bm{F}(\omega \pm i\eta) = P \int \frac{\mathrm{Im}A_F(\omega')d\omega'}{\omega - \omega'} \mp \pi \mathrm{Re}A_F(\omega),
\end{align}
 (similarly for impurity Green's function $G_{\mathrm{imp}}(z)$) and the self-energy is calculated as the matrix ratio:
\begin{equation}
\label{eq:selfEnergyTrick}
\bm{\Sigma}(z) = \bm{F}(z) \bm{G}_{\mathrm{imp}}(z)^{-1}.
\end{equation}
The calculation works well because the oscillatory NRG artifacts cancel out upon division, even for large values of the discretization parameter $\Lambda$.

After calculating the self-energy (both retarded and advanced), we calculate its spectral representation
\begin{equation}
\bm{A}_{\Sigma}(\omega) = \frac{i}{2\pi} \left(\bm{\Sigma}(\omega+i\eta) - \bm{\Sigma}(\omega-i\eta) \right).
\end{equation}

Due to numerical problems it may occur that spectral representation of the self-energy is not positive definite, $\bm{A}_\Sigma(\omega) < 0$. This violates the causality. If the positive definite requirement is broken at one iteration, the problem gets even more pronounced in the next one, leading to convergence problems. 
For each frequency, we thus make $\bm{A}_{\Sigma}(\omega)$ positive-definite. We find that the best procedure is to first make all diagonal elements positive, $A_{ii}>\epsilon_{\mathrm{clip}}$. We then use an eigenvalue decomposition, replace all negative eigenvalues with $\epsilon_{\mathrm{clip}}$ and reconstruct the matrix, thus enforcing it to be positive definite. We use a small clipping value $\epsilon_{\mathrm{clip} }=10^{-4}$. 

The self-energy has non-zero asymptotic value that has to be handled separately. We extract it from the high frequency asymptotic of $\bm{\Sigma}(\omega\pm i\eta)$ by fitting tails on left and right side with $\bm{B}^{\pm} + \bm{C}^{\pm}/z$, and then using $\bm{\Sigma}(\infty) = (\bm{B}^+ + \bm{B}^-)/2$. The reconstructed value is therefore
\begin{equation}
\label{eq:greenFromA}
\bm{\Sigma}^{\mathrm{clipped}}(\omega \pm i\eta) = \int \frac{\bm{A}_\Sigma^{\mathrm{clipped}}(\omega') d\omega'}{\omega \pm i\eta - \omega'} + \bm{\Sigma}(\infty).
\end{equation}

The CDMFT implemented on the real axis is very similar to that on the imaginary axis, with one difference: one needs to calculate both advanced and retarded version of all objects ($\bm{F}, \bm{G}_{\mathrm{imp}}, \bm{G},\bm{\Sigma}, \bm{\Delta}$) so we can go back to the spectral representation when the NRG is performed. In ideal case, $\bm{G}_{\mathrm{local} }(z)$ and $\bm{G}_{\mathrm{imp}}(z)$ would match in the converged solution. Due to the improved calculation of the self-energy in Eq. \eqref{eq:selfEnergyTrick},  $\bm{G}_{\mathrm{imp}}(z)$ has more broadening artifacts, so one should only match basic features. If the procedure converges to a non-physical solution, $\bm{G}_{\mathrm{imp}}(z)$ and $\bm{G}_{\mathrm{local}}(z)$ usually show very different features (for example, one has a gap, and the other does not). Restarting the calculation with a more appropriate initial $\bm{\Delta}(z)$ usually remedies the convergence problems in these cases. 

If not stated otherwise, the results are presented for the cubic lattice with half-bandwidth
$D$, therefore the hopping constant is $t/D=1/6$. 
We use the NRG discretization parameter $\Lambda=7$
and keep up to $N_{\mathrm{keep}}=4500$ states in each iteration in the CDMFT; this produces some energy
truncation artifacts, however much higher values of $N_{\mathrm{keep}}$ are not accessible due to memory constraints.
For consistency the same value of $\Lambda=7$ is used in the single-site DMFT as well. All calculations use 
the improved self-energy calculation, $N_z=8$ different interleaved meshes to reduce finite size
effects, and spectral broadening with the broadening parameter $b=0.35$. We use $5$ million $\bm{K}$ points with
artificial broadening $\eta=0.001$ to produce the local Green's function. The CDMFT calculations use
the discretization scheme described in the Appendix, while the DMFT calculations use the adaptive discretization without artifacts \cite{Zitko20091271}.

\section{Analysis of the systematic errors in the CDMFT(NRG)}
\label{sec:trunc}
There are two main parameters that control the artifacts of the impurity solver: the discretization parameter $\Lambda$ and
the maximum number of states kept after each iteration $N_{\mathrm{keep}}$. 

The NRG calculation for the two-site clusters have to be performed with high values of the discretization parameter $\Lambda$ and low $N_{\mathrm{keep}}$
in order to stay within memory constraints. The feasibility of
calculations with  a very high $\Lambda=7$ should be attributed to improved self-energy calculation,
without which the CDMFT produces very inaccurate results. Because the discretization grid is very coarse
at high $\Lambda$, it is important to use the $z$-averaging to reduce the finite-size effects.

\begin{figure}
\includegraphics[clip,width=0.49\textwidth]{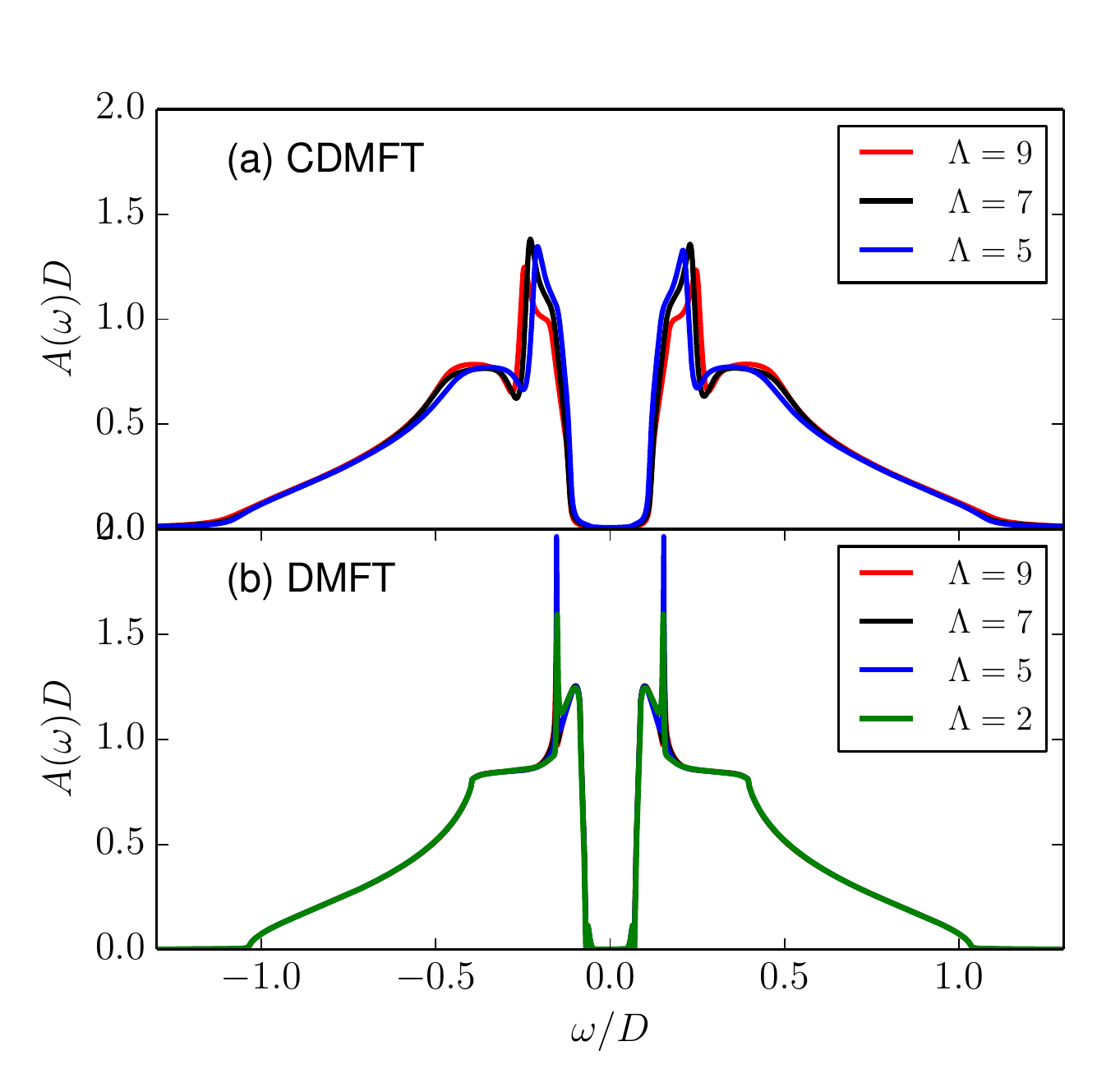}
\caption{Spin-averaged local spectral function for different values of the discretization parameter $\Lambda$ for the CDMFT (a) and DMFT (b).}
\label{fig:lambdaCompare}
\end{figure}

For $N_{\mathrm{keep}}=4500$, we see that lowering $\Lambda$
shifts peaks but does not change the signature features of the spectral function (Fig. \ref{fig:lambdaCompare}).
This is very encouraging because without high values of $\Lambda$,
the two-site cluster NRG computations are infeasible: for example, even
for $\Lambda=5$, one CDMFT iteration for a single interleaved mesh requires 
around $20$ hours on $16$ cores. The results are essentially unchanged in the DMFT (Fig. \ref{fig:lambdaCompare}b)
for a very broad range of $\Lambda$. Here we also plot $\Lambda=2$ result
that almost overlaps with other curves, thus we expect the DMFT results to be very close to the $\Lambda \rightarrow 1$ continuum limit even for surprisingly large $\Lambda$.  
Additionally, the AFM-KI phase transition point in the DMFT is not changed by $\Lambda$ (not shown).

The truncation plays a more important role than $\Lambda$ in the CDMFT. We describe the energy truncation at iteration $i$ with $E_{max}^i$: it is the energy of the state with the highest energy still not truncated. These energies for $J/D=0.3$ are presented in Table \ref{tab:cdmftTrunc}. Even for highest $N_{\mathrm{keep}}=8000$, the states in the first iteration are already truncated. We would need to keep $16000$ states to avoid truncation in the first iteration, and $16$ times more to avoid it in the second iteration. In the second iteration, we are omitting $94\%$ of the states, many in $[0,E_{\mathrm{cutoff}}]\Lambda^{-3/2}$ energy window. We would need to keep up to approximately $300000$ states to avoid truncation due to $N_{\mathrm{keep}}$ in the second iteration; the memory and computation requirements are too high for large values of $N_{\mathrm{keep}}$. The limit for $\sim 100 \mathrm{GB}$ memory is around $N_{\mathrm{keep}} = 8000$. 

\begin{table}
\begin{tabular}{| l || c | c | c | r | }
  \hline			
  $N_{\mathrm{keep}}$ & $E_{max}^1 / \Lambda^{-1}$ & $E_{max}^2 / \Lambda^{-3/2}$ & $E_{max}^{3}/\Lambda^{-2}$ & $E_{max}^{4}/\Lambda^{-5/2}$  \\
  \hline
  100 & 0.66 (0.6\%) & 1.64 & 3.12 & 3.6 \\
  500 & 0.718 (3\%) & 1.41 & 3.46 & 3.4 \\
  1500 & 1.23 (9.3\%) & 3.6 & 4.46 & 5.58 \\
  3000 & 1.74 (18.7\%) & 4.02 & 4.82 & 6.61 \\
  4500 & 1.79 (28\%) & 4.0 & 5.0 & 6.88 \\
  8000 & 2.34 (50\%) & 4.36 & 5.46 & 7.34 \\
  \hline  
\end{tabular}
 \caption{CDMFT truncation for $J/D=0.3$ and $z=1$. The percentage of states kept is shown in parenthesis; all states are already truncated in the first
 iteration. }\label{tab:cdmftTrunc}
\end{table}

\begin{figure} 
\includegraphics[clip,width=0.49\textwidth]{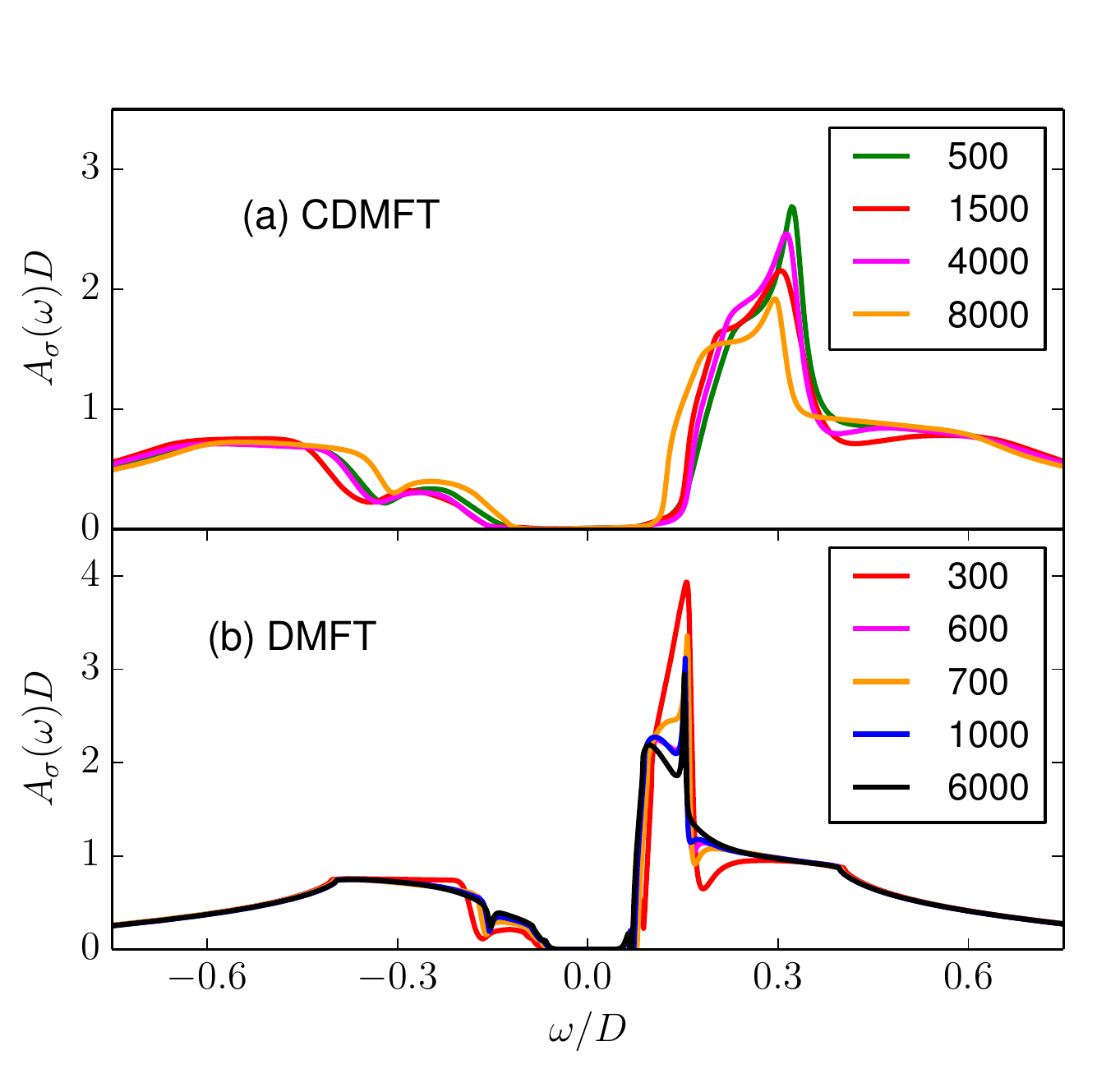}
\caption{Spin-resolved local spectral function for different $N_{\mathrm{keep}}$ in the CDMFT (a) and in the DMFT (b).}
\label{fig:Nkeep}
\end{figure}

In order to discuss the artifacts of the energy truncation, we study it more carefully in the DMFT where
we decrease $N_{\mathrm{keep}}$ to low values to emulate the effect of truncation in the CDMFT (Fig. \ref{fig:Nkeep}(b)). 
The results for $N_{\mathrm{keep}}=6000$ have no truncation artifacts at all and serve as the reference.
For $N_{\mathrm{keep}}=1000$, the curve still essentially overlaps with the reference. First difference appear for $N_{\mathrm{keep}}=700$,
where the curve around the spin resonance (peak left to the band edge) slightly broadens. The spin resonance becomes broader as we decrease $N_{\mathrm{keep}}$
down to $300$ but all other features remain at nearly the same position. As $N_{\mathrm{keep}}$ is decreased further,
the differences become qualitative; this is the limit where
we truncate too many states for the DMFT to converge to a stable and physically correct solution. In Table \ref{tab:dmftTrunc}, we show the highest energies of the states still taken into
account for first few iterations for interleaved mesh $z=1$. 
For $N_{\mathrm{keep}}>600$, the states are clipped by $E_{\mathrm{cutoff}}$ and we are reproducing all features, including the spin resonance.

Comparing the evolution of the spectral function for different value of $N_{\mathrm{keep}}$ in Fig. \ref{fig:Nkeep}(a), one can relate
the CDMFT solution for $N_{\mathrm{keep}}=500-8000$ to a region around $N_{\mathrm{keep}}\approx 300$ in the DMFT; this corresponds to spectral functions
that already resolve most of the features with the spin resonance broadened. If $N_{\mathrm{keep}}$ is too low in the CDMFT, the solution does not
converge. The high resolution region is not accessible to the CDMFT due to memory constraints.

This analysis indicates that even though we are use rather severe state truncation in the CDMFT, the results for $J/D=0.3$ and $T=0$ are qualitatively valid. We however expect quantitative deviations from ideal fully converged results. For instance, in order to capture phase transitions, the number of states required is increased. The CDMFT does not converge for $J/D=0.35-0.45$, where the phase transition between the AFM and the KI occurs. The DMFT emulation even with $N_{\mathrm{keep}}=700$ does not converge for $J/D=0.43$, and the area of non-convergence is wider if $N_{\mathrm{keep}}$ is decreased. The same is true for $T>0$ when capturing the thermal N\'eel transition. At higher temperatures, higher energy states become increasingly important as their contribution to the spectral function is increased. The effect of the state truncation is that the weight of the truncated states is transferred to the kept states. The kept states are used to describe the low-temperature behavior, thus the state truncation enhances the stability of the low-temperature phases. As a consequence, the N\'eel transition is shifted to higher temperatures. In the DMFT with $N_{\mathrm{keep}}=300$, for instance, the N\'eel temperature is raised by approximately a factor of $10$ compared to the fully converged results at high $N_{\mathrm{keep}}$. This explains the CDMFT artifacts when trying to capture the N\'eel transition, where the transition temperature is severely overestimated - in fact, we find the CDMFT transition temperature higher than in the fully converged  DMFT.

\begin{table}
\begin{center}
\begin{tabular}{| l || c | c | c | r | }
  \hline			
  $N_{\mathrm{keep}}$ & $E_{max}^1 / \Lambda^{-1}$ & $E_{max}^2 / \Lambda^{-3/2}$ & $E_{max}^{3}/\Lambda^{-2}$ & $E_{max}^{4}/\Lambda^{-2}$ \\
  \hline
  200 & 0.99 & 4.79 & 8.5 & 7.91 \\
  300 & 0.99 & 5.18 & 9.1 & 7.8 \\
  400 & 0.99  & 7.2 & 9.4 & 9.8 \\
  600 & 0.99 & 10 & 10 & 10 \\
  \hline  
\end{tabular}
 \caption{DMFT truncation for $J/D=0.3$ and interleaved mesh $z=1$.}\label{tab:dmftTrunc}
\end{center}
\end{table}

To test the accuracy further, we also calculated the results for the PAM, where we can compare them to the DMFT
and CDMFT calculated using the QMC as an impurity solver \cite{PhysRevB.84.115105}. For the N\'eel transition at
$V/D=0.18$, $\epsilon_f/D=-0.6$ and $U/D=1.2$, we obtain $T_N^{\mathrm{DMFT}}/D \approx 0.012$ that is very close to 
the QMC result, $T_N^{\mathrm{DMFT}}/D=0.015$. We however strongly overestimate the CDMFT transition that occurs at $T_N^{\mathrm{CMDFT}}/D=0.02$ in our case (for $N_{\mathrm{keep}}=4500$),
while using the QMC one finds $T_N^{\mathrm{CDMFT}}/D=0.004$. This further confirms that CDMFT(NRG) has severe artifacts for $T>0$
for accessible values of $N_{\mathrm{keep}}$. For this reason we focus on the $T=0$ case in the following.

\section{Fine structure of spectra in the Kondo lattice model}
\label{sec:main}

\begin{figure}
\includegraphics[clip,width=0.49\textwidth]{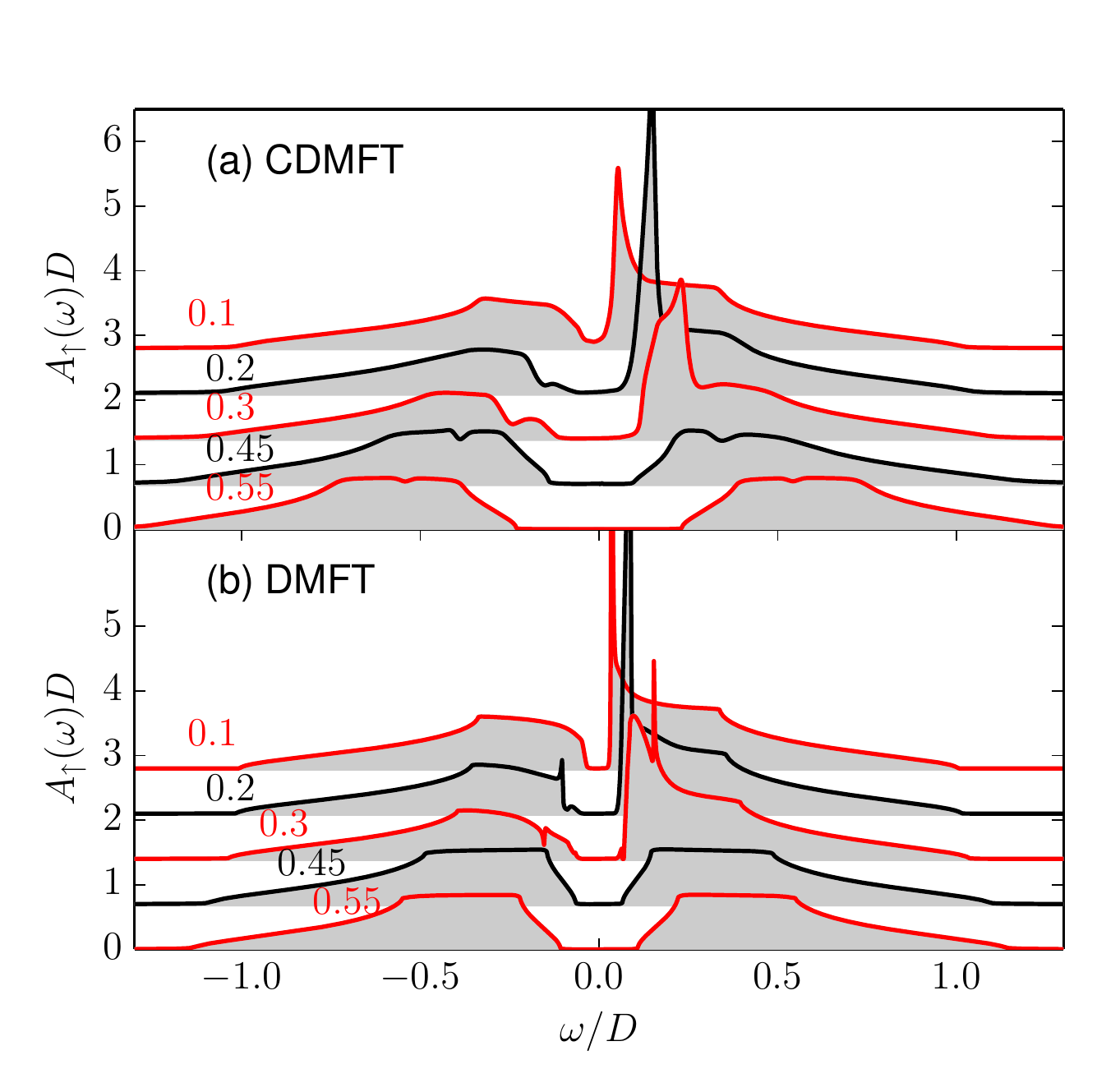}
\caption{Local spectral functions as $J$ is increased until a transition to a paramagnetic phase occurs.}
\label{fig:spinResonanceBuildUp}
\end{figure}

The origin of the antiferromagnetism at $T=0$ in the KLM depends on the value of the exchange coupling $J$. At low $J$, the system can be described by the Slater AFM (weak-coupling Hartree-Fock theory) \cite{rozenberg1995}. The main properties are inverse square root Slater singularities at the gap edges, with $c$ states only weakly polarized and $f$ states completely polarized in the opposite direction at the same site. The two-site CDMFT results nearly overlap with the DMFT results in this region, see Fig. \ref{fig:spinResonanceBuildUp} for $J/D=0.1$.

In the intermediate regime $J \sim 0.3D$, the correlations are stronger, and the system is no longer well described by weak-coupling theories. As $J$ is increased further, the Kondo mechanism prevails over the RKKY. The quantum phase transition between the AFM and the Kondo insulator within the CDMFT follows the same qualitative picture as in the single-site DMFT (Fig. \ref{fig:spinResonanceBuildUp}) but is shifted to lower values of $J$.

\begin{figure*}
\includegraphics[clip,width=0.95\textwidth]{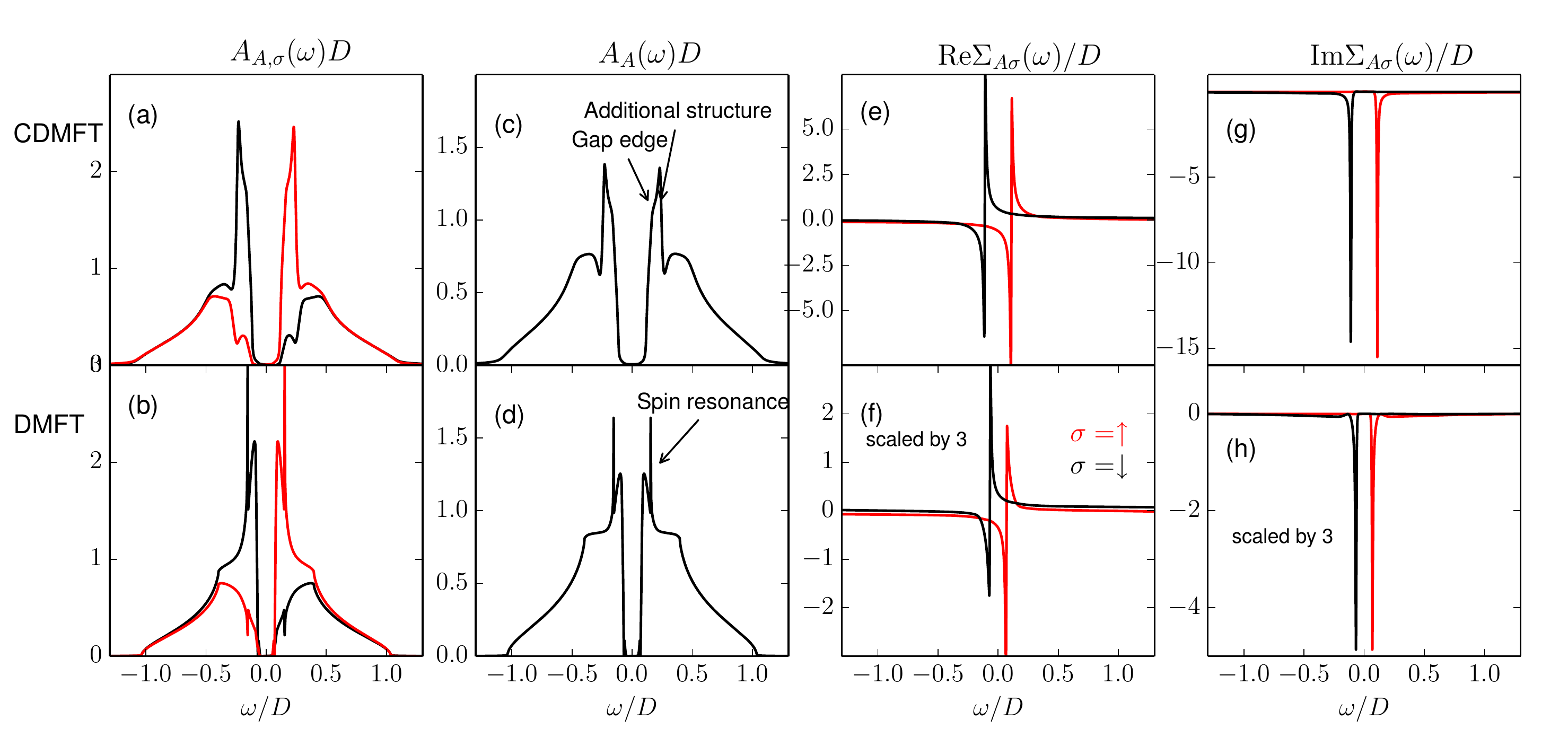}
\caption{Comparison of the CDMFT (top) and DMFT (bottom) results for $J/D=0.3$.}
\label{fig:resonanceMainFigure}
\end{figure*}

\begin{figure*}
\includegraphics[clip,width=0.95\textwidth]{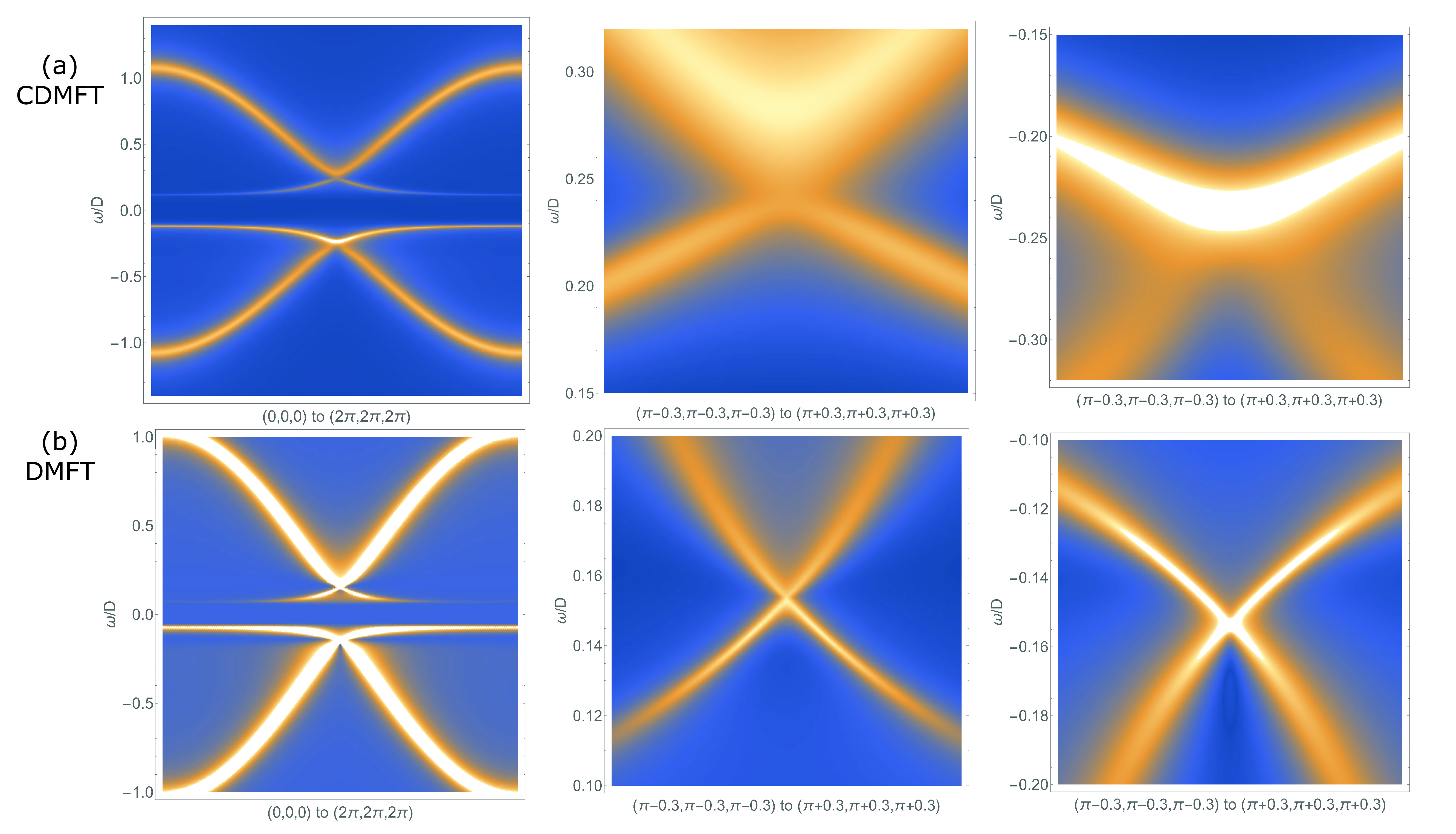}
\caption{(a) $\bm{K}$-resolved spectral functions for $J/D=0.3$ (strong coupling regime) within the CDMFT, with close-ups
to the region around $(\pi,\pi,\pi)$ responsible for the spin resonances. (b) $\bm{K}$-resolved spectral function $A_{A\uparrow}\left(\omega,\epsilon_{(k,k,k)} \right)$ for $J/D=0.3$ within the DMFT. }
\label{fig:epsRsolvedDMFT}
\end{figure*}

\begin{figure}
\includegraphics[clip,width=0.49\textwidth]{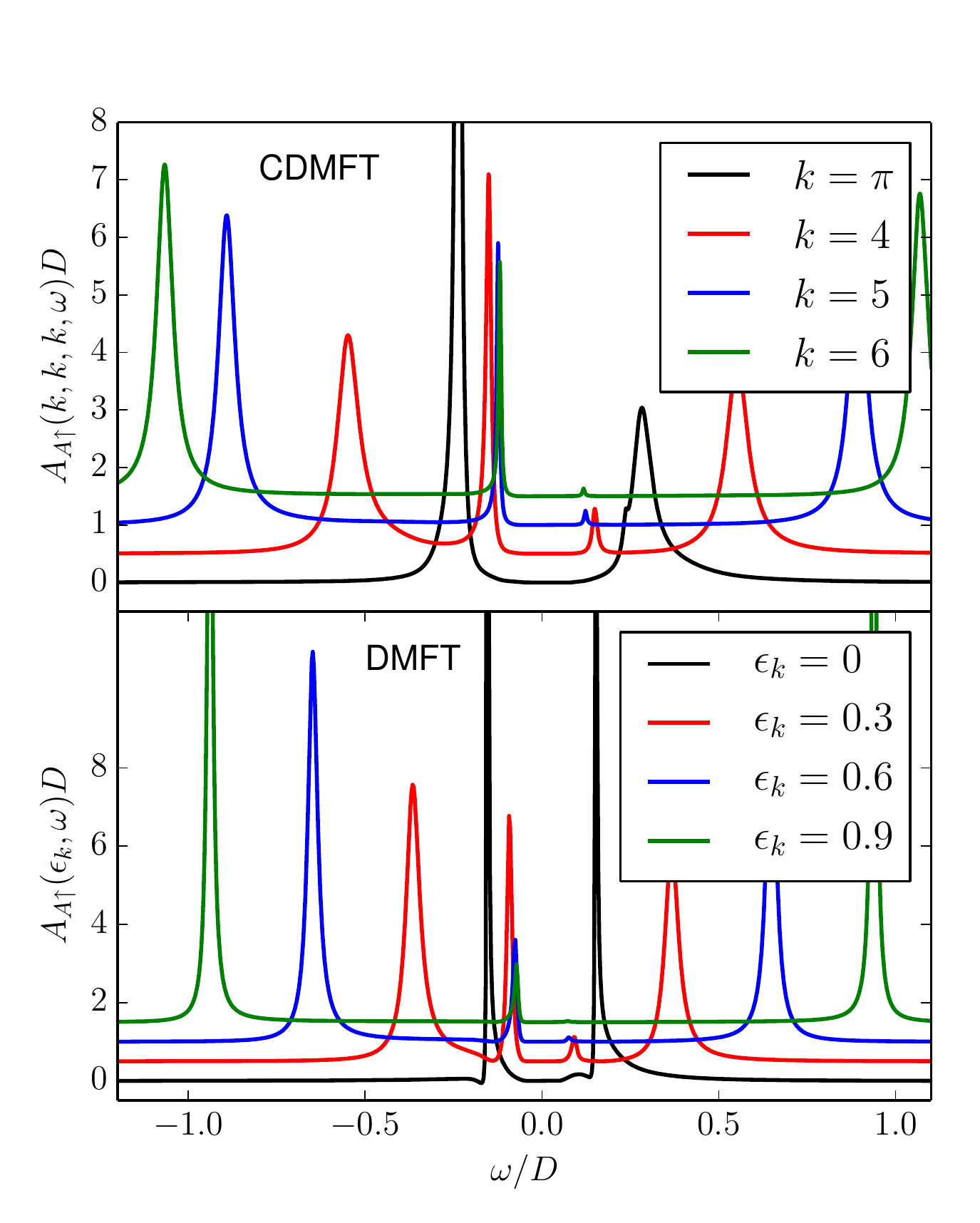}
\caption{(a) Spectral function at constant $\bm{K}$ within the CDMFT and (b) constant $\epsilon_K$ within the DMFT for $J/D=0.3$.}
\label{slicesInK}
\end{figure}

At low temperatures and for $J$ in the parameter range of strong Kondo effect, but still
in the antiferromagnetic phase, we find
an additional structure detached from band edge: the ``spin resonance''. We first discuss the DMFT results for $J/D=0.3$
presented in the second row in Fig.~\ref{fig:resonanceMainFigure}.
In the occupied band, there is a dip for minority
spin and a sharp peak for majority spin; the resonance is also visible
in the spin-averaged spectral function, $A = A_{\downarrow} +
A_{\uparrow}$ shown in Fig.~\ref{fig:resonanceMainFigure}(d).

The origin of these features can be traced to the momentum-resolved spectral function
$A(\vc{k},\omega)$, plotted in
Fig.~\ref{fig:epsRsolvedDMFT}(a). The close-ups on the regions where the
quasiparticle branches should intersect reveal that the spectral dip is
associated with a reduced spectral weight between the branches, i.e.,
an avoided crossing, while the peak corresponds to
an enhancement between two branches.

These features can be described in the extended hybridization picture in Eq. \eqref{eq:ansatz}
if we allow for the violation of quasi-local compensation ($H \neq -h$) and for the shifts of the poles
away from the real axis. The shape of the self-energy in
Fig.~\ref{fig:resonanceMainFigure}(f) is in at least qualitative agreement with the
one-pole structure, while the imaginary part in Fig.~\ref{fig:resonanceMainFigure}(h)
is nearly a $\delta$-function at the band edge, as expected. The DMFT results confirm the presence of additional
features around the energy $\omega^*$ defined in Eq. \eqref{eq:omegaStar}.

In the upper row in Fig. \ref{fig:resonanceMainFigure}, we show the corresponding results in the two-site CDMFT:
the spin resonances persist but become broader and obtain a Fano-resonance-like shape \cite{miroshnichenko2010fano}. The DMFT results
also indicate that the resonances have asymmetric shape but the resonance width is so small that we often do not see
the dip associated with the peak. The significant similarity between the CDMFT and the DMFT results is
also seen from the self-energies that have the same features but are different in
magnitude (the DMFT results in Fig. 
\ref{fig:resonanceMainFigure} for self-energies are scaled by $3$). This is directly related
to the increased gap in the CDMFT compared to the DMFT due to additional dynamic antiferromagnetic inter-site correlations
we take into account in the CDMFT; the same mechanism is responsible for the gap widening in the Hubbard model \cite{PhysRevB.76.045108}. 

The $\bm{K}$-resolved spectral function in the CDMFT (Fig. \ref{fig:epsRsolvedDMFT}(a)) shows
very similar structure to that in the DMFT, i.e. the spin resonances
are associated with the $(\pi,\pi,\pi)$ $\bm{K}$ point, corresponding to $\epsilon_K=0$ in the
DMFT. The slices of the $\bm{K}$-resolved spectral function (Fig. \ref{slicesInK})
emphasize the similarities in the structure with one difference: the CDMFT results
have broader peaks, especially for $\bm{K}=(\pi,\pi,\pi)$. There are two effects
that can contribute to broader peaks: interactions taken into account in the CDMFT exactly compared
to the mean-field treatment in the DMFT and the energy truncation artifacts in
the CDMFT. The truncation artifacts broaden all features and
are probably the main contribution to the decreased sharpness of the spin resonances.

\begin{figure} 
\includegraphics[clip,width=0.45\textwidth]{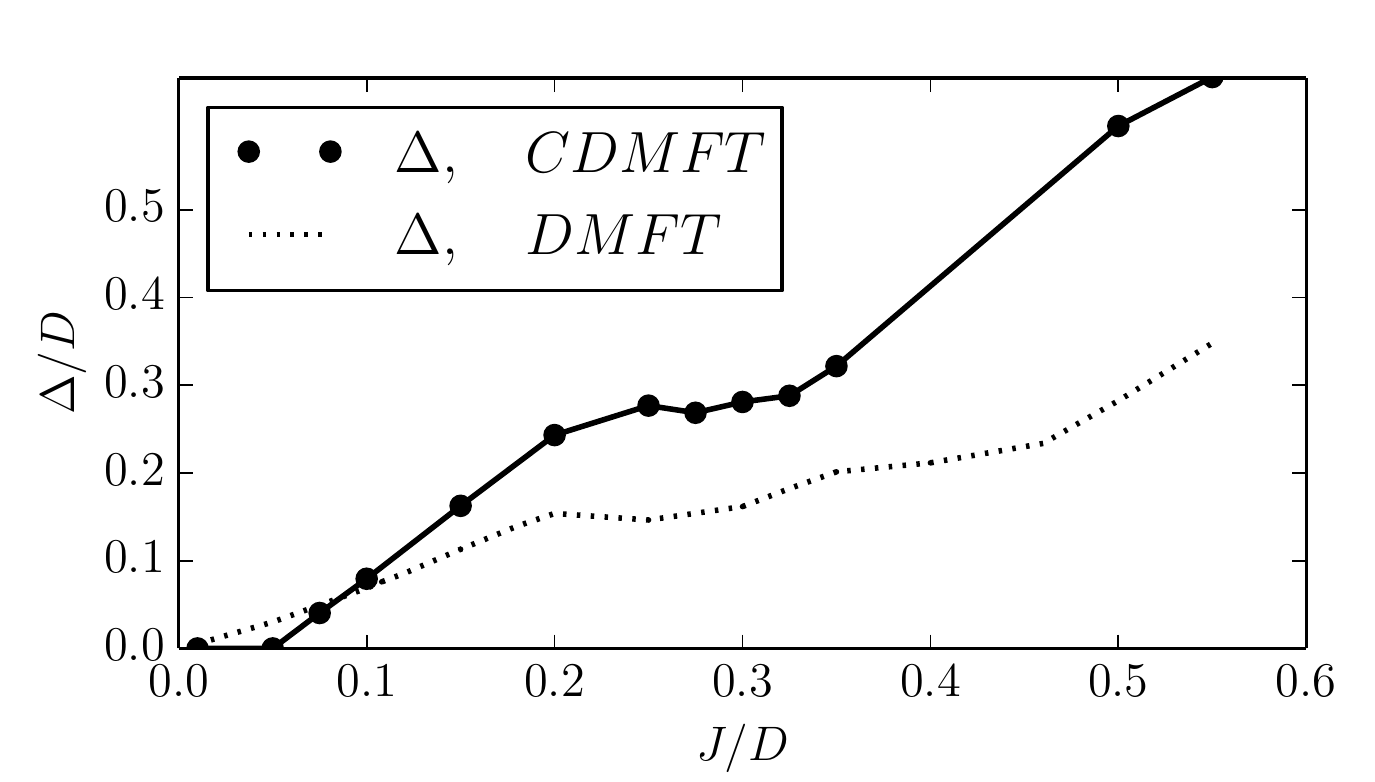}
\caption{Evolution of gap $\Delta$ as $J$ is increased with DMFT (dashed lines) and CDMFT (solid lines) for cubic lattice.}
\label{fig:allVsJ}
\end{figure}

The spin resonances appear around $J/D \approx 0.15$; at lower values, they are either not present or cannot be
distinguished from the band edge singularity. The build up of these resonances is shown in Fig. \ref{fig:spinResonanceBuildUp}. For
$J/D=0.2$, one can clearly see an additional feature for $\omega<0$ for both DMFT and CDMFT results while
the enhancement for $\omega>0$ is not yet detached from the band. Up to $J/D \approx 0.2$, the
gap $\Delta$ is still linear in $J$ in the DMFT results (dashed lines in Fig. \ref{fig:allVsJ}. In the CDMFT
 we cannot capture small gaps with the numerical $\bm{K}$-summing procedure.

\begin{figure} 
\includegraphics[clip,width=0.45\textwidth]{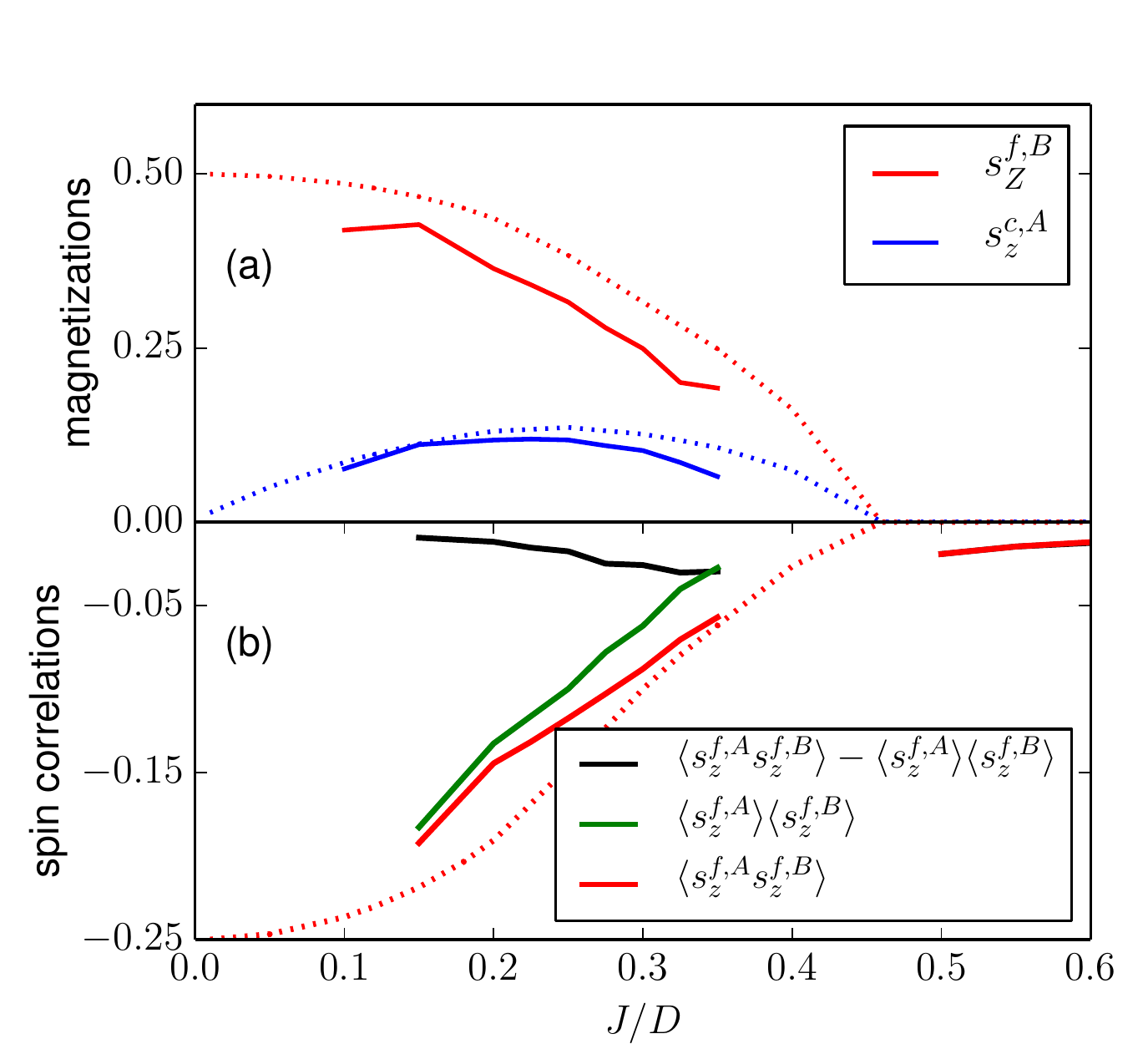}
\includegraphics[clip,width=0.45\textwidth]{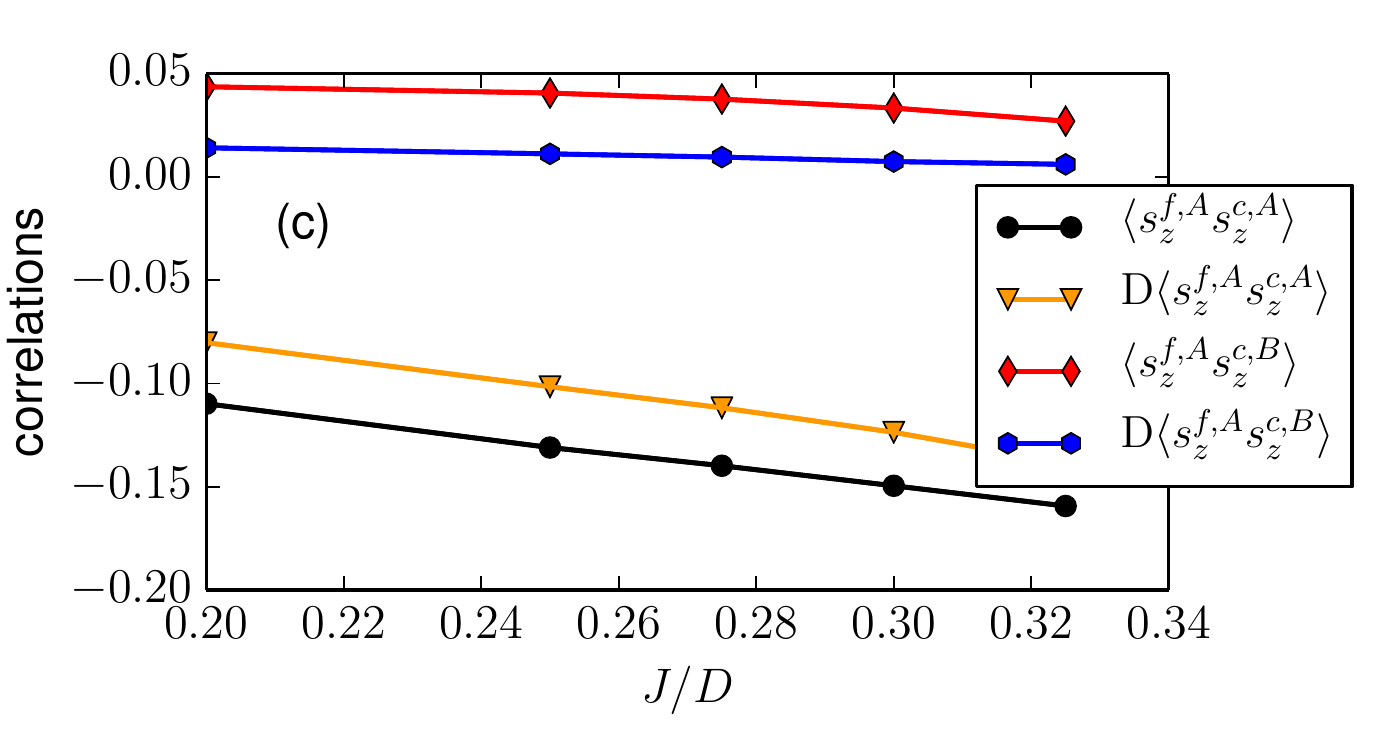}
\caption{(a) Sublattice magnetizations, (b) $f$ spin-spin correlations and (c) $f$ spin- $c$ spin correlations as $J$ is increased with CDMFT for cubic lattice. The operator $\mathrm{D}\langle AB \rangle$ is the dynamic part of the correlation, i.e. $\langle A B \rangle - \langle A \rangle \langle B \rangle$}
\label{fig:magVsJ}
\end{figure}

The $f$ spins are almost fully polarized for small $J$ in the DMFT, see Fig. \ref{fig:magVsJ}(a). The CDMFT suggest the same result
if we extrapolate the magnetization values to $J \rightarrow 0$, knowing that the result for $J/D=0.1$ already
has errors due to $K$-summing. The magnetization of the $c$ band goes to $0$ as $J \rightarrow 0$ in the DMFT and CDMFT, as expected for the
AFM \cite{rozenberg1995}. Overall, however, the stagerred magnetization of the CDMFT is smaller than in
the DMFT. The mean-field treatment usually overestimates the order, thus taking into account additional non-local processes decreases the order parameter.

Interestingly, the gap has a plateau for intermediate $J/D$ in the range $[0.2, 0.3]$, see Fig. \ref{fig:allVsJ}. The sublattice magnetization of the $c$ band increases up to $J/D \approx 0.2$ (and $0.25$ in the DMFT case), while the spin magnetization continuously decreases (Fig. \ref{fig:magVsJ}(a)).


The DMFT results indicate that the continuous phase transition from magnetic to paramagnetic phase occurs at $J^C_{\mathrm{DMFT}} = 0.45D$ (Fig. \ref{fig:magVsJ}): the magnetization continuously drops to zero, and the gap is continuous through the transition. The CDMFT has convergence difficulties for $0.35 < J/D < 0.45$, and we could not find a converged solutions in this region. We estimate that because the magnetization decreases faster than in the DMFT calculations, the phase transition occurs at lower $J$: our estimate is $J^C_{\mathrm{CDMFT}}=0.4D$.

For $J>J_C$, the system is a paramagnetic Kondo insulator. In Fig. \ref{fig:spinResonanceBuildUp}, one can see that all spin resonances disappear as soon as staggered order disappears. Interestingly, the gap in the Kondo insulator for the same $J$ is substantially larger than in the DMFT solution (Fig. \ref{fig:allVsJ}). Dynamic antiferromagnetic correlations between neighboring sites are large, and thus increase the gap size as in the AFM phase \cite{PhysRevB.76.045108}. 

\begin{figure} \centering
\includegraphics[clip,width=0.49\textwidth]{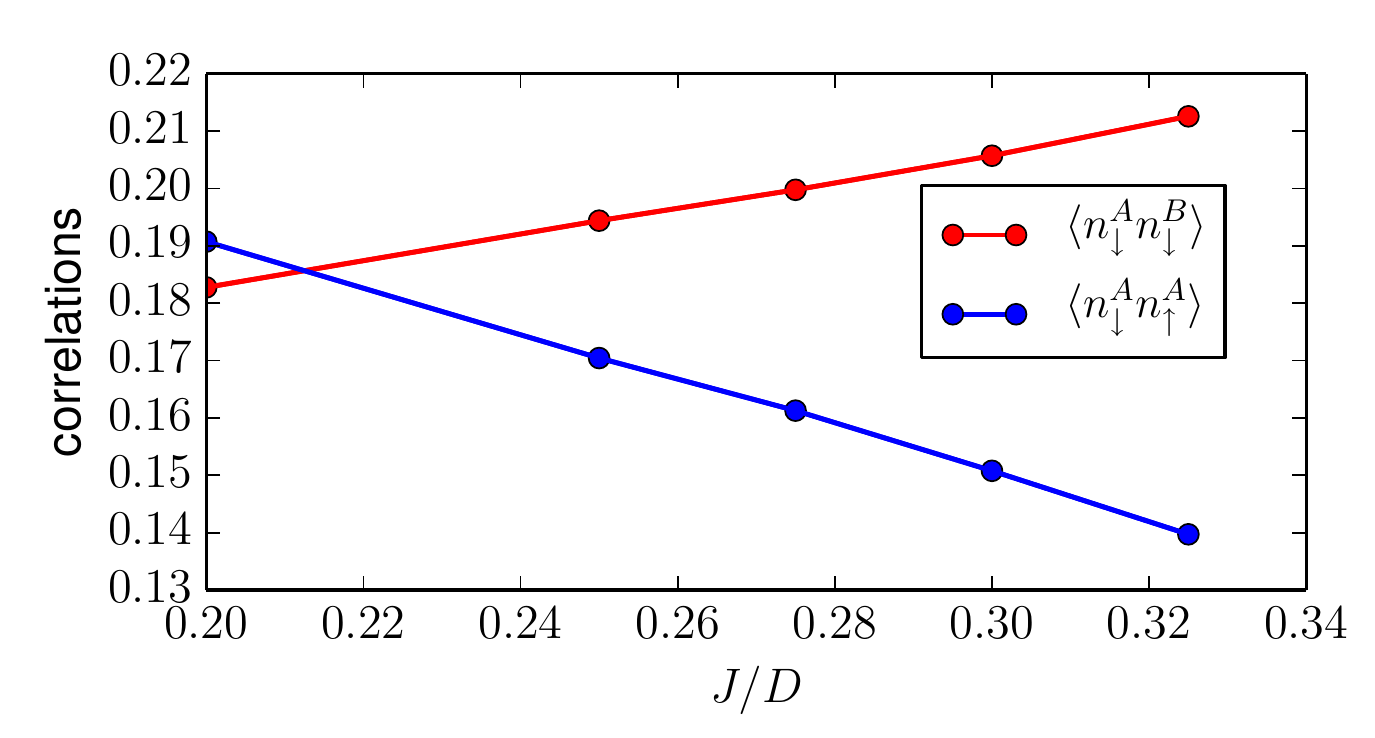}
\caption{Double occupancy and same spin occupancy for itinerant band ($c$) electrons on neighboring sites on the cubic lattice as a function of $J$. }
\label{fig:correlationsVsJ}
\end{figure}

Non-local correlations can be qualified through $\langle s_z^{f,A} s_z^{f,B} \rangle$,
i.e. what is the probability that the spin on $A$ aligns with spin on
sublattice $B$. In the antiferromagnetic solution at low $J$,
we expect the Slater AFM to be valid, with spins completely anti-aligned 
$\langle s_z^{f,A} s_z^{f,B} \rangle \rightarrow -1/4$. In the DMFT,
we are essentially treating correlation in the mean-field way,
and $\langle s_z^{f,A} s_z^{f,B} \rangle$ is equal to $\langle s_z^{f,A} \rangle \langle s_z^{f,B} \rangle$.
The DMFT results (Fig. \ref{fig:magVsJ}(b) with dashed lines)
start at the maximum value of $0.25$ for anti-aligned spins at low $J/D$, followed by a smooth transition to
no correlation in the paramagnetic solution.
In the CDMFT, however, there are additional dynamic contributions.
In Fig. \ref{fig:magVsJ}(b), we show that the dynamic part increases the
anti-alignment compared to the static value, however the total value 
is still lower in absolute value than in the DMFT, as expected. At $J/D=0.1$, there
are probably already artifacts in the CDMFT, so we cannot reliably
say that the correlation increases there. The dynamic
part increases in absolute value in the strongly
correlated regime around $J/D \approx 0.3$. The dynamic contribution is non-zero also in the paramagnetic solution for $J > J^{C}_{CDMFT}$,
where it falls slowly with increasing $J$. This is consistent with the CDMFT result
for the Hubbard model where nearest neighbor spin correlations are antiferromagnetic \cite{PhysRevB.76.045108}. In turn,
the short-range spin correlations are responsible for the gap widening because they enhance antiferromagnetism. The gap
is wider both in the AFM as in the Kondo insulator phase because the dynamic
correlations persist in the CDMFT solution through the phase transition. 

The double occupancy decreases slightly as $J$ is increased, see Fig. \ref{fig:correlationsVsJ}. At the same time, occupancy of same flavored spins at adjacent sites, $\langle n_\downarrow^A n_\downarrow^B \rangle$ increases with $J$.

One can also look at the correlations between the $c$ band and the $f$-moment on the same site 
(Fig. \ref{fig:magVsJ}(c)). 
The static (mean-field) value is very small as the magnetization of $c$
and $f$ sites decreases with increasing $J$, and the dynamic contribution promoting singlet formation dominates
, i.e. the Kondo effect is dominant. This
effect is already well described within the single-site DMFT framework. After the phase transition
only the dynamic contribution remains. The spin correlation between the $f$ spin
 on site $A$ and $c$ electron on site $B$, $\langle s_z^{f,A} s_z^{c,B} \rangle$, is
much smaller than the correlation on the same site. 

The hybridization ansatz from Eq. \eqref{eq:ansatz} can be extended to the
CDMFT case by including the matrix elements between the sites in the cluster, $\Sigma_{AB}(z)$. We thus write
\begin{equation}
\label{eq:NonLocalHybModel}
\bm{\Sigma}_{\uparrow}(\omega) = 
\begin{pmatrix}
h & t_{AB} \\
t_{AB} & - h
\end{pmatrix}
+
\tilde{V}^2 \left( z - \begin{pmatrix}
H & T_{AB} \\
T_{AB} & -H
\end{pmatrix}
+ i \delta \right)^{-1},
\end{equation}
where 
\begin{equation}
\bm{\Sigma}_{\uparrow}(\omega) = \begin{pmatrix}\Sigma_{AA,\uparrow}(\omega) & \Sigma_{AB,\uparrow}(\omega) \\ \Sigma_{BA,\uparrow}(\omega) & \Sigma_{BB,\uparrow}(\omega) \end{pmatrix}.
\end{equation}
New parameters $t_{AB}$ and $T_{AB}$ represent additional effective hoppings between $A$ and $B$ sites. The self-energy for spin $\downarrow$ is analogous with signs of $h$ and $H$ reversed, while the signs of $t_{AB}$ and $T_{AB}$ are unchanged.

\begin{figure} \centering
\includegraphics[clip,width=0.49\textwidth]{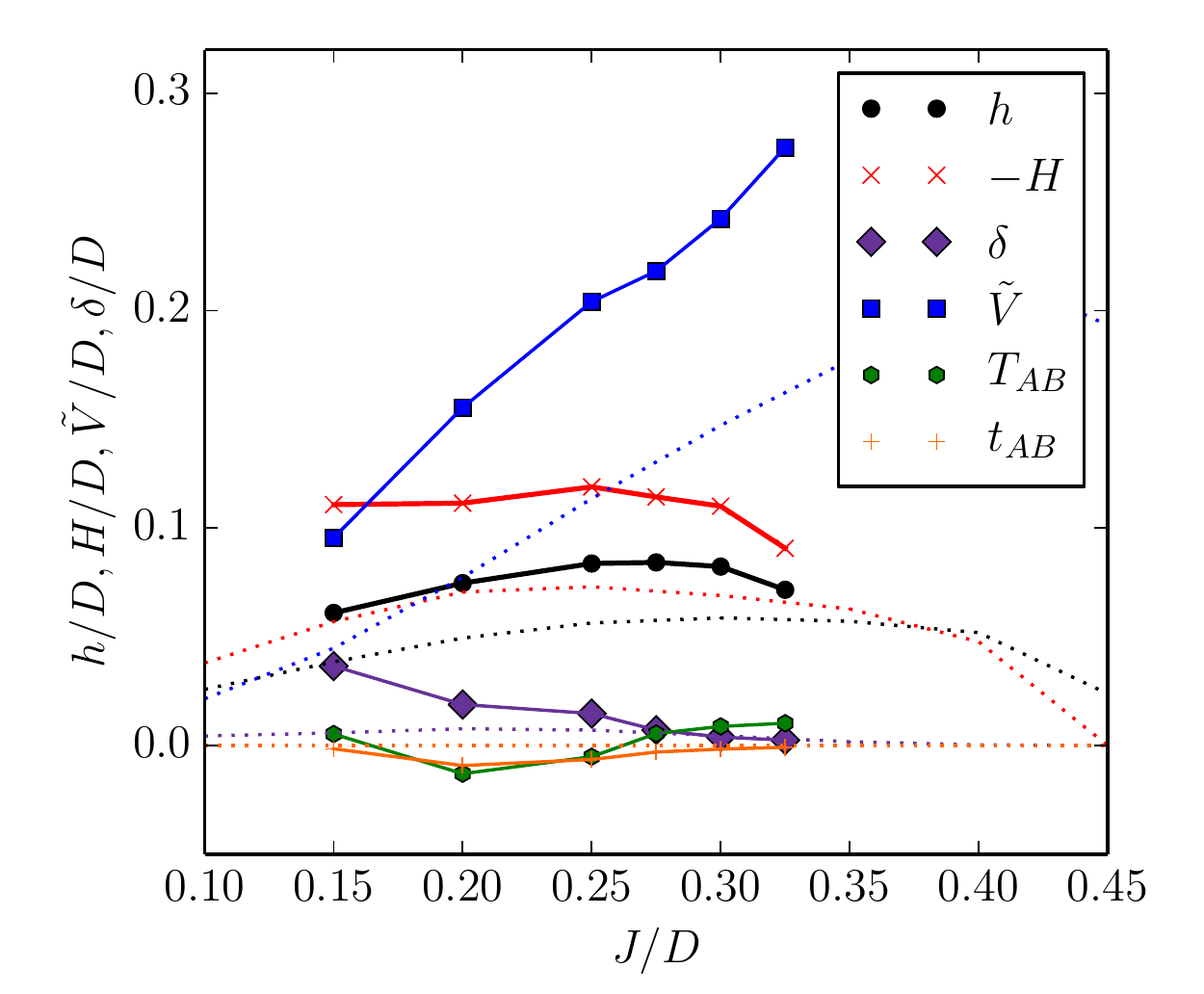}
\caption{Fit to Ansatz \eqref{eq:NonLocalHybModel} for CDMFT (solid lines) and DMFT (dashed lines) self-energies on
the cubic lattice.}
\label{fig:CDMFTModelFit}
\end{figure}

The parameters of the fit to the model are shown in Fig.\ref{fig:CDMFTModelFit}. The non-local correlations do not destroy the approximate validity of the hybridization picture,
as we are able to reproduce the self-energies for all parameters at least qualitatively. The fit is best in
the intermediate $J \approx 0.3$ regime and somewhat worse for small $J$, similarly to what is found in the DMFT \cite{PhysRevB.91.075105}. The difference
between $h$ and $-H$ in the CDMFT (Fig. \ref{fig:CDMFTModelFit}) confirms that the quasi-local compensation is violated, thus leading to new features around $\omega^* = \sqrt{\tilde{V}^2 + h^2}$. In fact, the difference
between $h$ and $-H$ is appreciably larger in the CDMFT compared to the DMFT (dashed lines). $\tilde{V}$ is substantially larger in the CDMFT compared to the DMFT for the same $J$, leading to larger gap in the CDMFT. The values of parameters $t_{AB}$ and $T_{AB}$ are very small.

\section{Conclusion}
We have performed a detailed study of the spectral properties of the KLM at half-filling using both the DMFT and the two-site cellular DMFT (CDMFT) to account for the non-local effects. We have confirmed the existance of fine structure (spin resonances) in the itinerant antiferromagnetic phase for $J<J_c$ also within the CDMFT. These features correspond to the bands at frequencies given by the crossing point of the quasiparticle branches in the center of the noninteracting band ($\pi,\pi,\pi$ point). They appear due to the inelastic-scattering processes which are not taken into account in the simplified (non-interacting)
hybridization picture. 

The short-range non-local correlations treated exactly with the CDMFT lower the value of the stagerred magnetization for both $f$ and $c$ sites compared to the DMFT, and thus also shift the quantum transition between the antiferromagnetic and Kondo insulator phase to lower value of $J_c$. The gap is wider in the CDMFT due to antiferromagnetic correlations between sites in the super-cell. These correlations persist through the phase transition and also widen the gap in the Kondo insulator phase.

The two-site CDMFT approach overestimates the paramagnet formation and thus suppresses the antiferromagnetic region too much \cite{PhysRevLett.95.237001}: $J_c$ for larger clusters would thus be somewhere between the two-site and the single-site DMFT result and the staggered magnetization at a specific $J$ would lie somewhere between the single-site DMFT and two-site CDMFT values.

This work  presents the first functional implementation of the CDMFT with the NRG as the impurity solver. The tests indicate that when using both the improved self-energy calculation to minimize the artifacts, as well as the interleaved meshes, one is able to perform the NRG despite very aggressive state truncations. One, however, needs to be careful when addressing regimes near phase transitions or when using it for finite temperatures, where state truncation leads to severe artifacts. The study of the antiferromagnetic phases away from half-filling is also not yet possible with NRG due to convergence problems. As possible future improvements, one could address those technical issues by either using more symmetries (especially for models that have less broken symmetries, like the paramagnetic Hubbard model) or using the interleaved NRG from Ref. \onlinecite{PhysRevB.93.235101}. Using larger supercells is probably prohibitive in the NRG, but one could use the non-local DMFT to study several two-site clusters and merge the correlations \cite{jabben2012extension}.

\section{ACKNOWLEDGMENTS}
R.\v{Z}. and \v{Z}.O. acknowledge the support of the Slovenian
Research Agency (ARRS) under Program No. P1-0044.

\appendix

\section{Discretization for multi-channel impurity problem}
\label{sec:appendix}
We show how to construct the coefficients for a generic impurity Hamiltonian without any symmetries:
\begin{equation}
\label{eq:generalImpurityProb}
H = H_{imp} + \sum_{kij} \tilde{\epsilon}_{kij} \tilde{c}_{k i}^\dagger \tilde{c}_{kj} + \sum_{kij} \left( \tilde{V}_{kij} d_{i}^\dagger \tilde{c}_{k j} +  \tilde{V}_{kij}^* \tilde{c}_{kj}^\dagger d_{i} \right),
\end{equation}
where $\tilde{\epsilon}_{kij}$ represent the on-site energies of the band operator $\tilde{c}$, while $\tilde{V}_{kij}$ describe the coupling of the impurity to the bath. The index $i=1,\dots,N_s$ in $f_i$ can represent spin, sub-lattice, orbital and/or any other degrees of freedom, while the index $j$ in $\tilde{c}_{kj}$ represents the bath degrees of freedom. The range of indices $i$ and $j$ need not be equal. The impurity Hamiltonian $H_{imp}$ can consist of any number of $d_i$ operators.


The first transformation is the diagonalization of $\tilde{\epsilon}_{kij}$ since for each $k$, the matrix $\bm{\tilde{\epsilon}}_k$ is Hermitian:
\begin{equation}
U_k \bm{\tilde{\epsilon}}_k U_k^\dagger = 
\begin{pmatrix}
\epsilon_{k1} & 0 & \cdots \\
0 & \epsilon_{k2} & \cdots \\
\cdots & \cdots & \cdots
\end{pmatrix}.
\end{equation}
Defining operators in a new basis, $\bm{c}_{k} \rightarrow U_k \tilde{\bm{c}}_{k}$ and $\bm{c}_{k}^\dagger \rightarrow \tilde{\bm{c}}_{k}^\dagger U_k^\dagger$, and using $V_{k} = \tilde{V}_{k} \cdot{} U_k$, we get
\begin{equation}
\label{eq:generalImpurityProb2}
H = H_{imp} + \sum_{ki} \epsilon_{ki} c_{k i}^\dagger c_{k i} + \sum_{kij} \left( V_{kij} d_{i}^\dagger c_{k j} +  V_{kij}^* c_{kj}^\dagger d_{i} \right).
\end{equation}
Hence the continuum Hamiltonian can always be transformed in a form with diagonal on-site terms, while the coupling term keeps full matrix dependency. We define the hybridization function in the matrix form,
\begin{equation}
\label{eq:generalHybridization}
\Gamma_{ij}(\omega) = \pi{}\sum_{k l} V_{kil} V_{kjl}^* \delta (\omega - \epsilon_{kl}).
\end{equation}
The hybridization describes how the electron hops from state $j$ on the impurity to bath and hops back on the impurity to state $i$, as we will see shortly in the derivation where we integrate out the bath degrees of freedom and this expression replaces the bath. Note, however, that the hybridization is a complex, positive definite quantity. We extend the function $\Gamma$ to the complex plane via
\begin{equation}
\label{eq:HybridizationOnZAxis}
\Delta_{ij}(z) = \int d\omega \frac{\Gamma_{ij}(\omega)}{z - \omega},
\end{equation}
where $z$ is a complex number off the real axis. To obtain back the representation on the real axis, one uses
\begin{equation}
\bm{\Gamma}(\omega) = \frac{i}{2} \left( \bm{\Delta}(\omega+i\eta) - \bm{\Delta}(\omega-i\eta) \right). 
\end{equation}

We would like to find a mapping to a continuous band Hamiltonian,
\begin{align}
\label{eq:generalGHModel}
H =& H_{imp} + \sum_{i} \int_{-1}^1 \mathrm{d}\epsilon g_i(\epsilon) a_{\epsilon i}^\dagger a_{\epsilon i} \nonumber \\
& + \sum_{ij} \int_{-1}^1 \mathrm{d}\epsilon \left( h_{ij}(\epsilon)  d_{i}^\dagger a_{\epsilon j} + h^*_{ji}(\epsilon) a_{\epsilon j}^\dagger d_{i} \right).
\end{align}
We  introduced a one-dimensional energy representation of states $a_{\epsilon i}$ with dispersion $g_i(\epsilon)$ and hopping to the impurity $h_{ij}(\epsilon)$. The energies are chosen to lie in the interval $[-1,1]$. We now relate the coefficients $\epsilon_{ki}$ and $V_{kij}$ (or the hybridization function $\Gamma$) to functions $g_i(\epsilon)$ and $h_{ij}(\epsilon)$.

The coefficients are not unique but a general relation that must hold can be derived using the action formalism. Because we deal with fermions, Grassmann algebra is required to the write the action. The partition function of the model \eqref{eq:generalImpurityProb2} is
\begin{equation}
Z_1 = \int_{imp} \int \Pi_{ki} D\psi_{ki}^\dagger D\psi_{ki} e^{S_1},
\end{equation}
where the integration goes over all impurity Grassmann variables ($\chi_{i}^+(\tau)$, $\chi_{i}(\tau)$ and other impurity variables), and band Grassmann variables $\psi_{ki}^+(\tau)$ and $\psi_{ki}(\tau)$. The action is given by
\begin{widetext}
\begin{align}
\label{eq:actionImpurityProb2Integral}
S_1(\chi, \chi^\dagger, \psi, \psi^\dagger) =& S_{imp}(\chi, \chi^\dagger, \cdots) 
+\int_0^{\beta} d\tau \sum_{ki} \left( \frac{\partial}{\partial \tau} - \epsilon_{ki} \right) \psi_{ki}^\dagger(\tau) \psi_{ki}(\tau^{-}) \nonumber \\
& -\int_0^{\beta} d\tau \sum_{kij}  \left( V_{kij} \chi_{i}^\dagger(\tau) \psi_{kj}(\tau^{-}) + V_{kij}^* \psi_{kj}^\dagger(\tau)\chi_i(\tau^{-})  \right).
\end{align}
\end{widetext}
We have replaced ordered operators with Grassmann variables in correct time slices ($c_{ki} \rightarrow \psi_{ki}(\tau^{-})$, $c_{ki}^\dagger \rightarrow \psi_{ki}(\tau)$). The $S_{imp}$ represents action on the isolated impurity that depends on $\chi$ and maybe some other Grassmann numbers, but not on $\psi$. The integration in Eq. \eqref{eq:actionImpurityProb2Integral} is actually representation for the sum over $N_\beta$ time slices. We are interested in the limit $N_{\beta} \rightarrow \infty$. We now explicitly write out the time slices, $\chi_{i}(\tau) \equiv \chi_{i, n}$ and $\chi_{i}(\tau^{-}) \equiv \chi_{i, n-1}$ (and the same for $\psi$),
\begin{widetext}
\begin{align}
\label{eq:actionImpurityProb2}
S_1(\chi, \chi^\dagger, \psi, \psi^\dagger) =& S_{imp}(\chi, \chi^\dagger, \cdots)  
-\frac{\beta}{N_{\beta}} \sum_{n=1}^{N_\beta} \sum_{ki} \epsilon_{ki} \psi_{ki,n}^\dagger \psi_{ki,n-1}  \nonumber \\ 
& -\frac{\beta}{N_{\beta}} \sum_{n=1}^{N_\beta} \sum_{kij} \left( V_{kij} \chi_{i,n}^\dagger \psi_{kj,n-1} + V_{kji}^* \psi_{kj,n}^\dagger \chi_{i,n-1} \right) 
+ \sum_{n=1}^{N_{\beta}} \sum_{ki} \psi_{ki,n}^\dagger ( \psi_{ki,n-1} - \psi_{ki,n} ).
\end{align}
\end{widetext}
The last term comes from the imaginary time derivative. We have to integrate over the bath degrees of freedom to obtain the action only concerning the impurity operators. Because the bath Grassmann variables only appear in the expression linearly and quadratically, we can use the Gaussian integral formula for Grassmann variables:
\begin{align}
\int D \psi^\dagger D\psi & \exp \left[ -\sum_{ij} \psi_i^\dagger G_{ij} \psi_j + \sum_i \psi_i \zeta_i^\dagger + \sum_i \psi_i^\dagger \zeta_i \right] = \nonumber \\ &= (\mathrm{det} G)^{-1/2} \exp \left[ -\sum_{ij} \zeta_i^\dagger (G^{-1})_{ij} \zeta_j \right].
\end{align}
Reshaping the action \eqref{eq:actionImpurityProb2} and integrating over all $\psi$, the effective action is
\begin{align}
\label{eq:effectiveActionImpurity}
S_{\mathrm{eff},1} =& S_{\mathrm{imp}}(\chi, \chi^\dagger, \cdots) + \nonumber \\
& \frac{\beta^2}{N_\beta^2} \sum_k \sum_{n,m=1}^{N_{\beta} } \sum_{ij,lp} V_{kil} V_{kjp}^* \chi_{l,n}^\dagger \chi_{p,m} (G_{1,k}^{-1})_{nl,mp},
\end{align}
with
\begin{equation}
G_{1,k} = 
\begin{pmatrix}
- \bm{I} & 0 & \cdots \\
\bm{I} - \beta{}/N_{\beta} \boldsymbol{\epsilon}_k\ & -\bm{I} & \cdots \\
\cdots & \cdots & \cdots
\end{pmatrix}.
\end{equation}
The first subscript of $G_{1,k}$ signals that $G$ is part of action $S_{\mathrm{eff},1}$, while the second index is the $k$-dependence; there is a different matrix for each $k$. The matrix indices are $(n,l)$ for rows and $(m,p)$ for columns; the $l,p$ indices are represented using the matrix (bold) notation inside the matrix $G_{1,k}$. Also, note that $\bm{\epsilon}_k$ is diagonal in
the internal space.

The derivation of the effective action for the Hamiltonian \eqref{eq:generalGHModel} is derived in a similar way
\begin{equation}
Z_2 = \int_{imp} \int \Pi_{ki} D \psi_{ki}^\dagger D\psi_{ki} e^{S_2},
\end{equation}
with
\begin{widetext}
\begin{align}
\label{eq:actionImpurityGHGeneral}
S_2(\chi, \chi^\dagger, \psi, \psi^\dagger) =& S_{imp}(\chi, \chi^\dagger, \cdots) 
-\frac{\beta}{N_\beta} \sum_{n=1}^{N_\beta} \sum_{i} \int d\mathrm{\epsilon} g_i(\epsilon) \psi_{i,n}^\dagger(\epsilon) \psi_{i,n-1}(\epsilon)  \nonumber \\ 
& -\frac{\beta}{N_\beta} \sum_{n=1}^{N_\beta} \sum_{ij} \int d\mathrm{\epsilon} \left( h_{ij}(\epsilon) \chi_{i,n}^\dagger \psi_{j,n-1}(\epsilon) + h_{ji}^*(\epsilon) \psi_{j,n}^\dagger (\epsilon ) \chi_{i,n-1} \right) 
+ \sum_{n=1}^{N_\beta} \sum_{i} \int d\mathrm{\epsilon} \psi_{i,n}(\epsilon) ( \psi_{i,n-1}(\epsilon) - \psi_{i,n}(\epsilon) ).
\end{align}
\end{widetext}
The effective action is
\begin{align}
\label{eq:effectiveActionImpurityGM}
&S_{\mathrm{eff},2} = S_{\mathrm{imp} }(\chi, \chi^\dagger, \cdots) + \nonumber \\ 
&+\frac{\beta^2}{N^2} \int d\mathrm{\epsilon} \sum_{n,m=1}^{N_\beta} \sum_{ij,lp} h_{il}(\epsilon) h_{jp}^*(\epsilon) \chi_{l,n}^\dagger \chi_{p,m} (G_{2}^{-1}(\epsilon))_{nl,mp},
\end{align}
with
\begin{equation}
G_{2}(\epsilon) = 
\begin{pmatrix}
- \bm{I} & 0 & \cdots \\
\bm{I} - \beta{}/N \bm{g}(\epsilon) & -\bm{I} & \cdots \\
\cdots & \cdots & \cdots
\end{pmatrix},
\end{equation}
where $\bm{g}(\epsilon)$ is a diagonal matrix.
To compare actions, we first use that $G_{1}$ and $G_{2}$ are diagonal in $l,p$ index, and we can rewrite the sum over $l,p$ to only sum over $l$ in both effective actions \eqref{eq:effectiveActionImpurity} and \eqref{eq:effectiveActionImpurityGM}. We can also rewrite the sum over $k$ in \eqref{eq:effectiveActionImpurity} using the relation
\begin{align}
\sum_k V_{ilk} V_{jlk}^* (G_{1,k})^{-1}_{nl, ml} = \nonumber \\ \int d\mathrm{\epsilon} \sum_k V_{ilk}V_{jlk}^* \delta(\epsilon - \epsilon_{kl}) (G_{1,\epsilon})^{-1}_{nl, ml}.
\end{align}
Because $G_{1,k}$ only depends on $k$ through $\epsilon_k$, we can change the notation to $G_{1}(\epsilon_k) \equiv G_{1,k}$. We now require $\left( G_{1}(g_l(\epsilon) \right)^{-1} = \left( G_{2}(\epsilon) \right)^{-1}$: there might be other possibilities but we can fix this condition and try to find a correspondence of actions in this form. If we want the action to be the same, all the coefficients must be the same for every $i,j,n,m,\epsilon$, and we get
\begin{align}
\int d\epsilon \sum_l h_{il}(\epsilon) h_{il}^*(\epsilon) (G_1 (g_l(\epsilon)) )^{-1}_{nl,ml} = \nonumber \\ \int d\epsilon \sum_{l,k} V_{ilk}V_{jlk}^* \delta(\epsilon - \epsilon_{kl}) (G_{1}(\epsilon))^{-1}_{nl, ml},
\end{align}
where $g_l$ and $h_{il}$ are functions to be determined. Changing the order of summation and integration, and for each sum, changing the integration variable of the left side, $x_l=g_l(\epsilon)$, one obtains
\begin{align}
\sum_l \int d\mathrm{x}_l & \frac{\partial f_l(x_l)}{\partial x_l} h_{il}(f_l(x_l) ) h_{jl}^*(f_l(x_l)) (G_1 (x_l ))^{-1}_{nl,ml} = \nonumber \\  &= \int d\mathrm{x} \sum_{l,k} V_{ilk}V_{jlk}^* \delta(x - \epsilon_{kl}) (G_{1}(x))^{-1}_{nl, ml},
\end{align}
with $f_l(x)$ the inverse of $g_l(x)$.

Comparing both sides, we arrive at the equation that must hold for all $i$, $j$ and $x$:
\begin{equation}
\sum_l \frac{\partial f_l(x)}{\partial x} h_{il}(f_l(x)) h_{jl}^*(f_l(x)) = \frac{1}{\pi} \Gamma_{ij}(x),
\end{equation}
where we have used the definition of the hybridization in Eq. \eqref{eq:generalHybridization} for the right part. We are now free to choose $h_{il}(x)$ and $f_l(x)$ in any way to satisfy this equation.

The simplest scheme is to choose a linear dispersion in the continuous model, $g_l(x)=x$, meaning that $f_l(x)=x$. We can use an eigenvalue decomposition
\begin{equation}
\bm{h} (x) \bm{h}^\dagger(x) = \frac{1}{\pi} \bm{\Gamma}(x) = ( \bm{U} \frac{1}{\sqrt{\pi}} \sqrt{ \bm{D} } ) (\sqrt{ \bm{D}} \frac{1}{\sqrt{\pi}} \bm{U}^\dagger)
\end{equation}
where $\bm{D}$ is diagonal with nonegative elements because $\bm{\Gamma}$ is positive definite, $\bm{\Gamma} > 0$, therefore we take the square root of $\bm{D}$ element-wise. $\bm{\Gamma} > 0$ follows from the definition in Eq. \eqref{eq:generalHybridization}; it is constructed by $\bm{V} \bm{V}^\dagger$ that is positive definite by construction for each $k$, and because each part of the sum is positive definite, the final expression is also positive definite. We can therefore express $\bm{h}(x)$ as
\begin{equation}
\label{eq:GammaHRelation}
\bm{h} (x) = \frac{1}{\sqrt{\pi}} \sqrt{ \bm{\Gamma}(x)}.
\end{equation}
We now perform logarithmic discretization of energies $x^{\pm}_n = \pm{} \Lambda^{1-n-z}$, where the corresponding intervals $I_n^{\pm}$ are $[-x_n, -x_{n+1}]$ and $[x_{n+1}, x_n]$. The $\Lambda$ paramater is typically around $2$ for high precision single band calculations but it can be as high as $8$ and still give reliable results, especially for static properties. Decreasing $\Lambda$ increases the computation cost exponentially \cite{PhysRevB.79.085106,PhysRevB.49.11986}. 

In each interval, we only allow constant wave function created by an operator $a_{n \sigma \alpha}^\dagger$. We set $h_{ij}$ by averaging over the interval, and obtain
\begin{align}
\left( \sum_l h_{il}^{\pm,n} h_{jl}^{\pm, *n} \right) =& \frac{1}{d_n}\int^{\pm, n} d\mathrm{\epsilon} \left( \sum_l h_{il}^{\pm}(\epsilon) h_{jl}^{\pm, *}(\epsilon) \right) d\mathrm{\epsilon} = \nonumber \\ 
&= \frac{1}{\pi{} d_n}\int^{\pm,n} d\mathrm{\epsilon} \Gamma_{ij}(\epsilon)
\end{align}
where we have just taken the average value of $h$ in each of the integrals. Using the relation \eqref{eq:GammaHRelation}, we can express it using the hybridization,
\begin{equation}
\label{eq:h_n_constant}
\bm{h}_n^\pm = \frac{1}{\sqrt{\pi{} d_n}} \sqrt{\int^{\pm,n} d\mathrm{\epsilon} \bm{\Gamma}(\epsilon) }.
\end{equation}
The approximation made in this discretization scheme is that we neglect all non-constant wave functions at each interval ($p \neq 0$). In most NRG discretization schemes this step is justified by noting that the impurity does not couple to those states, and that they are therefore a correction in the next order. In our scheme, the impurity states couple also to $p \neq 0$ states that we simply drop out of the calculation, so we can expect further artifacts compared to other schemes. These artifacts, however, still go to zero in the continuum limit $\Lambda \rightarrow 1$.

We can now rewrite the discretized conduction energy part of Eq. \eqref{eq:generalGHModel} on a specific interval as
\begin{equation}
\int^{\pm{}, n} d\mathrm{\epsilon} g_{i}(\epsilon) \rightarrow \int^{\pm{},n} \epsilon \cdot d\mathrm{\epsilon},
\end{equation}
Combining all the above expression, we get a logarithmically discretized Hamiltonian
\begin{align}
\label{eq:NRGTridiag}
H = H_{imp} + \sum_{n i} \left( \xi_{n i }^+ a_{n i}^\dagger a_{n i} + \xi_{n i }^- b_{n i}^\dagger b_{n i} \right) + \nonumber \\
\frac{1}{\sqrt{\pi}} \left[ \sum_{ij} d_i^\dagger \sum_n \left( \gamma_{nij}^+ a_{nj} + \gamma_{nij}^- b_{nj} \right) + h.c. \right]
\end{align}
with
\begin{equation}
\gamma_{nij}^\pm = \left( \int^{\pm, n} d\mathrm{x} \bm{\Gamma}(x) \right)^{1/2}_{ij}
\end{equation}
and
\begin{equation}
\xi_{n, i}^\pm = \frac{1}{d_n}\int^{\pm, n} x \cdot{} dx.
\end{equation}

Note that linear dispersion is not the only possible way to discretize. A more sophisticated approach is to first set $\bm{h}$ to a constant value in each interval (step functions) and the try to find the best possible candidate for $g_l(\epsilon)$. In this scheme, the impurity couples directly only to kept states. We formulated this method as a least square minimization with the constraint that $g_l(\epsilon)$ is bijective but have had limited success. The least square minimization sometimes converges to a local minimum, and the overall coefficients do not always respect the symmetries of the hybridization. At the time of writing the article, an alternative approach to discretization in eigenvector basis, possibly with less artifacts, has been presented in Ref. \onlinecite{PhysRevB.93.035102}. 

\section{Wilson chain for multi-channel impurity problem}
\label{sec:wilsonChain2}

We introduce matrix notation for operators $\bm{a}_{n}$,
\begin{equation}
\bm{a}_{n} = \begin{pmatrix}a_{n,1} \\ a_{n,2} \\ \cdots \\ a_{n,N} \end{pmatrix} \;\; \text{and} \;\; \bm{a}_n^\dagger = \left( a_{n,1}^\dagger, a_{n,2}^\dagger, ..., a_{n,N}^\dagger \right),
\end{equation}
and similarly for $\bm{b}$ and semi-infinite chain operators $\bm{f}$. They all obey the anti-commutator relations
\begin{equation}
\left\{  \bm{a}_{n}, \bm{a}_{n}^\dagger \right\} = \bm{1}.
\end{equation}
Using new definitions, we can rewrite the Hamiltonian in Eq. \eqref{eq:NRGTridiag} in the matrix form, 
\begin{align}
\label{eq:generalNxNDiscProblem}
H = H_{imp} +& \sum_n \left(\bm{a}_n^\dagger \bm{\xi}_n^+ \bm{a}_n + \bm{b}_n^\dagger \bm{\xi}_n^- \bm{b}_n  \right) \nonumber \\ &+ \frac{1}{\sqrt{\pi} } \sum_n \left( \bm{d}^\dagger \bm{\gamma}_n^+ \bm{a}_n + \bm{d}^\dagger \bm{\gamma}_n^- \bm{b}_n + \right. \nonumber \\
& \left. + \bm{a}_n^\dagger (\bm{\gamma}_n^{+})^\dagger \bm{d} + \bm{b}_n^\dagger  (\bm{\gamma}_n^{-})^\dagger \bm{d} \right).
\end{align}
Note that $\bm{\gamma}_n^\pm$ and $\bm{\xi}_n^\pm$ are all Hermitian $N \times N$ matrices. Known derivations deal with real matrices but they can be complex as well: magnetic field or any cluster computation have complex coefficients.

We now define the semi-infinite chain using a transformation
\begin{align}
\label{eq:generalNXNUV}
\bm{f}_n &= \sum_m \left( \bm{U}_{nm} \bm{a}_m + \bm{V}_{nm} \bm{b}_m \right),  \nonumber \\
\bm{a}_n &= \sum_m \bm{U}^\dagger_{mn} \bm{f}_m, \nonumber \\
\bm{b}_n &= \sum_m \bm{V}^\dagger_{mn} \bm{f}_m, 
\end{align}
where $f_n$ are vectors with $N$ components. In the matrix form, we can write the above expression as
\begin{equation}
\bm{f} = \begin{pmatrix} \bm{U} & \bm{V} \end{pmatrix} \cdot{} \begin{pmatrix} \bm{a} \\ \bm{b} \end{pmatrix}.
\end{equation}
There is an orthogonality relation
\begin{equation}
\label{eq:generalNxNOrtho}
\sum_n \left( \bm{U}_{mn}\bm{U}_{mn}^\dagger + \bm{V}_{mn}\bm{V}_{mn}^\dagger \right) = \bm{I}
\end{equation}
present for each $m$ because they represent transformation to a new basis and the normalization of the wavefunction must be preserved. The matrices $\bm{U}_{nm}$ and $\bm{V}_{nm}$ depend on $4$ indices; two for the internal space: $i,j=1,...N$, an index that is connected to the number of $f$ states: $m=0,1,2,...,N_{f}$ represents index in the chain, and an index that represents the interval in the star Hamiltonian, $n=0,1,...,N_{int}$. The Hermitian $\mathbf{U}^\dagger$ conjugation is performed only on the internal space.
The impurity couples directly to $\mathbf{f}_0$. It is therefore equal to (from Eq. \eqref{eq:generalNxNDiscProblem}),
\begin{equation}
\mathbf{f}_0 = \left( \sqrt{\bm{\zeta}} \right)^{-1} \sum_n \left( \bm{\gamma}^+_n \bm{a}_n + \bm{\gamma}^-_n \bm{b}_n \right),
\end{equation}
where
\begin{equation}
\bm{\zeta} = \sum_n \left( \bm{\gamma}^+_n \cdot{}  \bm{\gamma}^+_n  + \bm{\gamma}^-_n \cdot{} \bm{\gamma}^-_n  \right).
\end{equation}
We can take the matrix square root of $\bm{\zeta}$ because it is positive definite and invertible. The coefficients of unitary transformation are
\begin{align}
\bm{U}_{0n} &= \left( \sqrt{\bm{\zeta}} \right)^{-1} \bm{\gamma}^+_n, & \bm{V}_{0n} &=  \left( \sqrt{\bm{\zeta}} \right)^{-1} \bm{\gamma}^-_n .
\end{align}
This transformation is unitary (first row of matrix coefficients sums to an identity matrix) due to proper normalization. Together with the transformation of the conduction band using Eqs. \eqref{eq:generalNXNUV}, the Hamiltonian has the form
\begin{align}
\label{eq:generalNxNFinalDisc}
H = H_{imp} + &\sum_{n,m,l} \left( \bm{f}_{l}^\dagger \bm{U}_{ln} \bm{\xi}^+_{n} \bm{U}^\dagger_{mn} \bm{f}_m + \bm{f}_{l}^\dagger \bm{V}_{ln} \bm{\xi}^-_{n} \bm{V}^\dagger_{mn} \bm{f}_m \right) + \nonumber \\
&+ \sqrt{\frac{\bm{\zeta}} {\pi{}}} \left( \bm{d}^\dagger \bm{f}_0 + \bm{f}_0^\dagger \bm{d} \right).
\end{align}
We are looking for a transformation to the semi-infinite chain form, explicitly written as
\begin{align}
\label{eq:generalNxNchain}
H = H_{imp} + &\sum_n \left( \bm{f}_n^\dagger \bm{\epsilon}_n \bm{f}_n + \bm{f}_n^\dagger \bm{t}_n \bm{f}_{n+1} + \bm{f}_{n+1}^\dagger \bm{t}^\dagger_n \bm{f}_n \right) + \nonumber \\
&+ \sqrt{\frac{\bm{\zeta}} {\pi{}}} \left( \bm{d}^\dagger \bm{f}_0 + \bm{f}_0^\dagger \bm{d} \right).
\end{align}
We can now compare these two equations. We treat the $\mathbf{f_0}$ operator separately because it is at the end of the chain. The comparison gives us the equality
\begin{equation}
\label{eq:generalNXNfull}
\sum_{n,l} \left( \bm{f}_{l}^\dagger \bm{U}_{ln} \bm{\xi}^+_{n} \bm{U}^\dagger_{0n}  + \bm{f}_{l}^\dagger \bm{V}_{ln} \bm{\xi}^-_{n} \bm{V}^\dagger_{0n} \right) = \bm{f}_0^\dagger \bm{\epsilon_0} + \bm{f}_1^\dagger \bm{t}_0^\dagger.
\end{equation}
By taking the left anti-commutator $\{ \bm{f}_0, \cdot{} \}$ and using the relation $\{ \bm{f}, \bm{f}^\dagger \bm{A} \} = \bm{A}$, we obtain
\begin{equation}
\bm{\epsilon}_0 = \sum_n \left( \bm{U}_{0n} \bm{\xi}^+_n \bm{U}^\dagger_{0n} + \bm{V}_{0n} \bm{\xi}^-_n \bm{V}^\dagger_{0n} \right).
\end{equation}
We can now express $\bm{f}_0^\dagger \bm{\epsilon}_0$ with operators $\bm{a}_n$ and $\bm{b}_n$. In Eq. \eqref{eq:generalNXNfull}, we can rewrite $\bm{f}_l^\dagger$ operators on the left hand side back to $\bm{a}_n$ and $\bm{b}_n$, insert the expression of $\bm{f}_0^\dagger \bm{\epsilon}_0$ and get
\begin{align}
\label{eq:generalNxNf_1}
\bm{f}_1^\dagger \bm{t}^\dagger_0 = \sum_n &\left( \bm{a}_n^\dagger \left( \bm{\xi}_n^+ \bm{U}^\dagger_{0n} - \bm{U}^\dagger_{0n}\bm{\epsilon}_0 \right) + \right. \nonumber \\
& \left.  \bm{b}_n^\dagger \left( \bm{\xi}_n^- \bm{V}^\dagger_{0n} - \bm{V}^\dagger_{0n}\bm{\epsilon}_0 \right) \right).
\end{align}
By anti-commutation of the above expression with the Hermitian conjugate of itself, and using the relation $\left\{ \bm{B} \bm{f}, \bm{f}^\dagger \bm{A} \right\} = \bm{B}\bm{A}$, where $\bm{A}, \bm{B}$ are matrices, $\bm{t}_0$ is obtained:
\begin{align}
\label{eq:generalNxNt0}
\bm{t}_0\bm{t}^\dagger_0 =& \sum_n \left[ \left( \bm{U}_{0n} \bm{\xi}_n^{+\dagger} - \bm{\epsilon}_0^\dagger \bm{U}_{0n} \right) \left( \bm{\xi}_n^+ \bm{U}^\dagger_{0n} - \bm{U}_{0n}^\dagger \bm{\epsilon}_0 \right) + \right. \nonumber \\ &\left. + \left( \bm{V}_{0n} \bm{\xi}_n^{-,\dagger} - \bm{\epsilon}_0^\dagger \bm{V}_{0n} \right) \left( \bm{\xi}_n^- \bm{V}_{0n}^\dagger - \bm{V}_{0n}^\dagger\bm{\epsilon}_0 \right) \right]
\end{align}
All the matrices entering in the above equation are Hermitian, therefore the result is also a Hermitian matrix. The hopping matrix $\bm{t}_0$ is positive definite, so we just take the matrix square root to obtain it. 

Expanding the left side of Eq. \eqref{eq:generalNxNf_1} into operators $a_n$ and $b_n$ and comparing the coefficients, the $\bm{U}_{1n}$ and $\bm{V}_{1n}$ are expressed as
\begin{align}
\bm{U}_{1n}^\dagger = \left( \bm{\xi}^{+}_{n} \bm{U}_{0n}^\dagger - \bm{U}_{0n}^\dagger \bm{\epsilon}_0 \right) \left( \bm{t}_0^{\dagger} \right)^{-1}, \nonumber \\
\bm{V}_{1n}^\dagger = \left( \bm{\xi}^{-}_{n} \bm{V}_{0n}^\dagger - \bm{V}_{0n}^\dagger \bm{\epsilon}_0 \right) \left( \bm{t}_0^{\dagger} \right)^{-1}.
\end{align}
With the same procedure, we now treat operators $\bm{f}_m$, $m>0$. Comparing the semi-infinite chain Hamiltonian in Eq. \eqref{eq:generalNxNchain} and Eq. \eqref{eq:generalNxNFinalDisc} yields
\begin{align}
\label{eq:generalNxNmg0}
\sum_{n,l} & \left( \bm{f}_{l}^\dagger \bm{U}_{ln} \bm{\xi}^+_{n} \bm{U}^\dagger_{mn}  + \bm{f}_{l}^\dagger \bm{V}_{ln} \bm{\xi}^-_{n} \bm{V}^\dagger_{mn} \right) = \nonumber \\
&\bm{f}_m^\dagger \bm{\epsilon_m} + \bm{f}_{m-1}^\dagger \bm{t}_{m-1} + \bm{f}_{m+1}^\dagger \bm{t}_m^\dagger.
\end{align}
Again taking the anti-commutator $\{ \bm{f}_m^\dagger, \cdot{} \}$, the on-site energy matrix is obtained
\begin{equation}
\bm{\epsilon}_m = \sum_n \left( \bm{U}_{mn} \bm{\xi}_n^+ \bm{U}_{mn} + \bm{V}_{mn} \bm{\xi}_n^- \bm{V}_{mn} \right).
\end{equation}
We proceed by writing the Eq. \eqref{eq:generalNxNmg0} in term of operators $\bm{a}_n, \bm{b}_n$,
\begin{align}
\sum_n &\left( \bm{a}_n^\dagger \bm{\xi}_n^+ \bm{U}^\dagger_{mn} +  \bm{b}_n^\dagger \bm{\xi}_n^- \bm{V}^\dagger_{mn} \right) = \sum_n \left( \bm{U}_{mn}^\dagger \bm{a}_n^\dagger + \bm{V}_{mn} \bm{b}_n^\dagger \right) \bm{\epsilon_m} \nonumber \\
&+ \sum_n \left( \bm{U}_{m+1,n}^\dagger \bm{a}_n^\dagger + \bm{V}_{m+1,n} \bm{b}_n^\dagger \right) \bm{t}_m^\dagger \nonumber \\
&+ \sum_n \left( \bm{U}_{m-1,n}^\dagger \bm{a}_n^\dagger + \bm{V}_{m-1,n} \bm{b}_n^\dagger \right) \bm{t}_{m-1} .
\end{align}
Expressing $\bm{f}_m^\dagger \bm{t}_m^\dagger$ and calculating the anti-commutator with its Hermitian form allows us to calculate $\bm{t}_m$ (one additional term compared to Eq. \eqref{eq:generalNxNt0}),
\begin{widetext}
\begin{align}
\bm{t}_m\bm{t}_m^\dagger = \sum_n & \left[ \left( \bm{U}_{mn} \bm{\xi}_n^{+\dagger} - \bm{\epsilon}_0^\dagger \bm{U}_{mn} - \bm{t}_{m-1}^\dagger \bm{U}_{m-1,n} \right) \left( \bm{\xi}_n^+ \bm{U}_{mn}^\dagger - \bm{U}_{mn}^\dagger\bm{\epsilon}_m - \bm{U}^\dagger_{m-1,n} \bm{t}_{m-1} \right) + \right. \nonumber \\ & 
\left.
\left( \bm{V}_{mn} \bm{\xi}_n^{-,\dagger} - \bm{\epsilon}_0^\dagger \bm{V}_{mn} - \bm{t}_{m-1}^\dagger \bm{V}_{m-1,n}  \right) \left( \bm{\xi}_n^- \bm{V}_{mn}^\dagger - \bm{V}_{mn}^\dagger\bm{\epsilon}_m - \bm{V}^\dagger_{m-1,n} \bm{t}_{m-1} \right) \right]
\end{align} 
\end{widetext}
New coefficients are expressed as
\begin{align}
\bm{U}_{m+1,n}^\dagger = \left( \bm{\xi}_n^+ \bm{U}_{mn}^\dagger - \bm{U}_{mn}^\dagger\bm{\epsilon}_m - \bm{U}_{m-1,n}^\dagger\bm{t}_{m-1} \right) \left( \bm{t}_m^{\dagger} \right)^{-1}, \nonumber \\
\bm{V}_{m+1,n}^\dagger = \left( \bm{\xi}_n^- \bm{V}_{mn}^\dagger - \bm{V}_{mn}^\dagger\bm{\epsilon}_m - \bm{V}_{m-1,n}^\dagger\bm{t}_{m-1} \right) \left( \bm{t}_m^{\dagger} \right)^{-1}.
\end{align}
A set of recursive relations for the orthogonal transformations has been thus derived. In the implementation, it is important to calculate using high precision (using arbitrary precision arithmetic). A good value is $3000$ mantissa bits. We also force normalization (unitary) of the transformation, Eq. \eqref{eq:generalNxNOrtho}) after each iteration. Another important strategy is to enforce the symmetries, especially for bigger matrices. We use a mask of zeros and ones to enforce zero entries in the  matrices $\bm{\xi}_n$ for the parts that should be $0$. This is important because small errors can amplify exponentially in Wilson chain iterative procedure. It is important to check that no such symmetry breaking occurs for the chain lengths considered. As a rule of thumb, one can simulate chain lengths of around $65$ sites for $4$ active complex coefficients (for example full $2 \times 2$ matrices) and around $25$ for $8$ coefficients (for example $4 \times 4$ matrices with the half of the elements set to $0$ due to symmetry). This seems to be sufficient in practice because we do not need many coefficients for high $\Lambda$.

\bibliography{KLM_CDMFT}

\end{document}